\documentclass[intlimits,twoside,a4paper]{article}

\usepackage{amsmath,amssymb}
\usepackage{graphicx}

\usepackage[cp1251]{inputenc}

\usepackage[eqsecnum]{cmpj3}

\issue{2019}{22}{1}{13703}
\doinumber{10.5488/CMP.22.13703}

\title[Investigation on phosphorene nanosheet]%
{Nitrogen dioxide and ammonia gas molecules interaction studies on phosphorene nanosheet --- a~DFT investigation }%

\author[V. Nagarajan, R. Chandiramouli]{V. Nagarajan, R. Chandiramouli\thanks{Corresponding author}}
\address{
School of Electrical and Electronics Engineering, Shanmugha Arts Science Technology and Research Academy (SASTRA) University, Tirumalaisamudram, Thanjavur, Tamil nadu --- 613 401, India 
}

\date{Received October 16, 2018, in final form February 24, 2019}

\begin{document}

\maketitle

\begin{abstract}
The adsorption behaviour of hazardous gas molecules, namely nitrogen dioxide (NO$_2$) and ammonia (NH$_3$), on phosphorene nanosheet (PNS) was explored by means of ab initio technique. To improve the structural solidity of pristine PNS, we have introduced the passivation of hydrogen and fluorine at the terminated edge. The structural solidity of both hydrogen and fluorine passivated PNS is verified in terms of formation energy. The main objective of this research work is to probe NO$_2$ and NH$_3$ gases using PNS as a base sensing material. The adsorption of various preferential adsorption sites of these gas molecules is studied in accordance with the average HOMO-LUMO gap changes, natural-bond-orbital (NBO) charge transfer, HOMO-LUMO gap, and adsorption energy. Notably, the negative value of adsorption energy is found upon the adsorption of NO$_2$ and NH$_3$ on PNS and it is in the range of $-1.36$ to $-2.45$~eV. The findings of the present research work recommend that the hydrogenated and fluorinated PNS can be effectively used as a chemical sensor against NO$_2$ and NH$_3$ molecules.

\pacs 71.15.Mb 
\end{abstract}

\section{Introduction}

Detection of toxic gas molecules is the most important and required criterion especially regarding industrial chemical processing, preparation of drugs, public safety, agriculture, environmental monitoring, and indoor air quality control. A few decades ago, metal oxide semiconductors (MOS) played a crucial role in industrial process control and environmental monitoring \cite{1}. MOS is the most extensively used gas sensing material owing to its fast redox reaction mechanism, low cost, high sensitivity towards different hazardous gases, simplicity in measurements with low power consumption and high compatibility \cite{2,3,4}. Long recovery periods, low specificity of MOS-based gas sensors limit the sensitivity and measurement accuracy \cite{5}. However, these drawbacks in MOS-based gas sensors are overcome with new sensor technology, which utilizes nanowire, nanotube, and two-dimensional nanosheets. Recently, two-dimensional (2D) nanosheets built with a few or a single atomic layer, such as metal dichalcogenides (for instance, MoS$_2$ and MoSe$_2$) and graphene-based nanomaterials, represent a promising class of materials and attract considerable interest owing to its physicochemical properties and unique structures \cite{6,7,8,9}, which are related to abundant active sites and large surface areas. These properties provide 2D nanosheets with exciting scenarios for various applications in sensors, electronic devices, energy conversion/storage devices and catalysis \cite{10,11,12}. Moreover, the previous reports on these types of materials have inspired chemists, physicists, and engineers to further investigate novel two-dimensional nanosheets such as oxides \cite{13}, carbides \cite{14}, nitrides \cite{15} and phosphates \cite{16}. 

Black phosphorus (BP) material is one of the stable allotropes of phosphorus in which the atomic layers are organized through van der Waals (vdW) interactions. Furthermore, the geometry of BP is the same as graphite; the elemental substance of black phosphorus can be exfoliated into an ultrathin nanosheet. Even though phosphorus belongs to group V element, bulk BP material is a p-type semiconductor with the energy band gap of 0.3~eV \cite{17,18,19}. Recently, many researchers have reported that the energy gap of the bulk form of BP can be modified from 0.3~eV to 2.0~eV, for monolayer BP. The properties of BP mainly depend on the number of layers under 2D limit \cite{17,18}. 

Phosphorene has attracted great interest among the scientific community owing to its exciting properties, which can be synthesized by sticky-tape technique \cite{18,19}. It exhibits (1) natural surface passivation excluding any dangling bond; (2) quantum confinement; (3) strong interaction with light; (4) no lattice mismatch issues and (5) high surface area \cite{20,21,22}.  These properties are suitable for the application of photocatalysis. Moreover, phosphorene shows anisotropic properties, and tunable bandgap leads to high mobility quantum transport, batteries, gas storage, and infrared optoelectronics applications \cite{23}. 	In recent times, Kaloni et al. \cite{24} have reported the dissociation of water molecules in nanoseconds using group IV monochalcogenides. Kou et al. \cite{25} investigated the interaction behaviour of various gases on a single layer phosphorene with its $I-V$ characteristics.  Lalitha et al. \cite{26} reported calcium doped and calcium decorated phosphorene as the base material for adsorption of CO$_2$, CH$_4$, H$_2$, and NH$_3$ molecules. Ray et al. \cite{27} explored the gas sensing characteristics of 2D single crystals such as phosphorene, graphene, and MoS$_2$. Cui et al. \cite{28} studied the layer-based sensing performance and sensitivity of NO$_2$ gas sensor on phosphorene. Anurag Srivastava and co-workers \cite{29} proposed the electron transport properties and sensitivity of NO$_2$ and NH$_3$ gas molecule on black phosphorene along with $I-V$ characteristics and electronic properties. We demonstrated the interaction behaviour of NO$_2$ molecules on armchair PNS molecular system for chemical sensor application using density functional theory (DFT) studies \cite{30}. Based on these aspects, we conducted the literature survey with the support of the SCOPUS database. After conducting the literature survey, we realized that there are limited reports with reference to the DFT technique to explore the interaction properties of NO$_2$ and NH$_3$ gas molecules on two-dimensional H-PNS and F-PNS. The inspiration behind the suggested work is to explore NO$_2$ and NH$_3$ interaction properties on H-PNS and F-PNS and to identify the most suitable interaction site. In the present research work, passivated hydrogen and fluorine PNS have been utilized as a base substrate for probing the NO$_2$ and NH$_3$ molecules.

\section {Quantum chemical calculations}

The hydrogenated and fluorinated PNS are relaxed and simulated in accordance with Gaussian 09 package \cite{31}. The interaction of NO$_2$ and NH$_3$ gas molecules on PNS is also examined through this package. In this work, hybrid Becke’s three (B3) and correlation functional Lee-Yang-Parr (LYP) with 6-31g(d) basis set is applied \cite{32,33,34}. To investigate the interaction behaviour of NO$_2$ and NH$_3$ gas molecules on PNS, the selection of prominent and suitable basis set is one of the important measures. Anurag Srivastava et al. \cite{29} proposed the interaction behaviour of phosphorene using B3LYP functional with 6-31g(d) basis set. Further, to make a calculation more precise, we included B3LYP-D3, a hybrid GGA functional \cite{35}. In addition to B3LYP-D3 functional, two more functionals namely,  PBE0-D3 and B97-D3 are included in the latest Grimme’s dispersion term --- D3 version (available in Gaussian 09) \cite{36}. Using the dispersion-corrected DFT \cite{37,38}, we calculated the adsorption energy of PNS upon interaction of NO$_2$ and NH$_3$ molecules. Hydrogen and fluorine termination is carried out to eliminate the edge effect. Moreover, B3LYP-D3/6-31g(d) provides good results for H-PNS and F-PNS with pseudopotential approximation. The choice of the basis set for phosphorene is also validated with the published work of Anurag Srivastava et al. \cite{29}. Using Gauss Sum 3.0 utility, we have calculated the DOS-spectrum, lowest unoccupied (LU-) and highest occupied (HO-) molecular orbital of PNS \cite{39}. The energy convergence for H-PNS and F-PNS is observed in the range of $10^{-5}$~eV while investigating the interaction properties of NO$_2$ and NH$_3$ on PNS.
\newpage

\section {Results and discussion}

\subsection {Geometric structural details of phosphorene nanosheet}

\begin{figure}[!b]
\centerline{\includegraphics[width=0.6\textwidth]{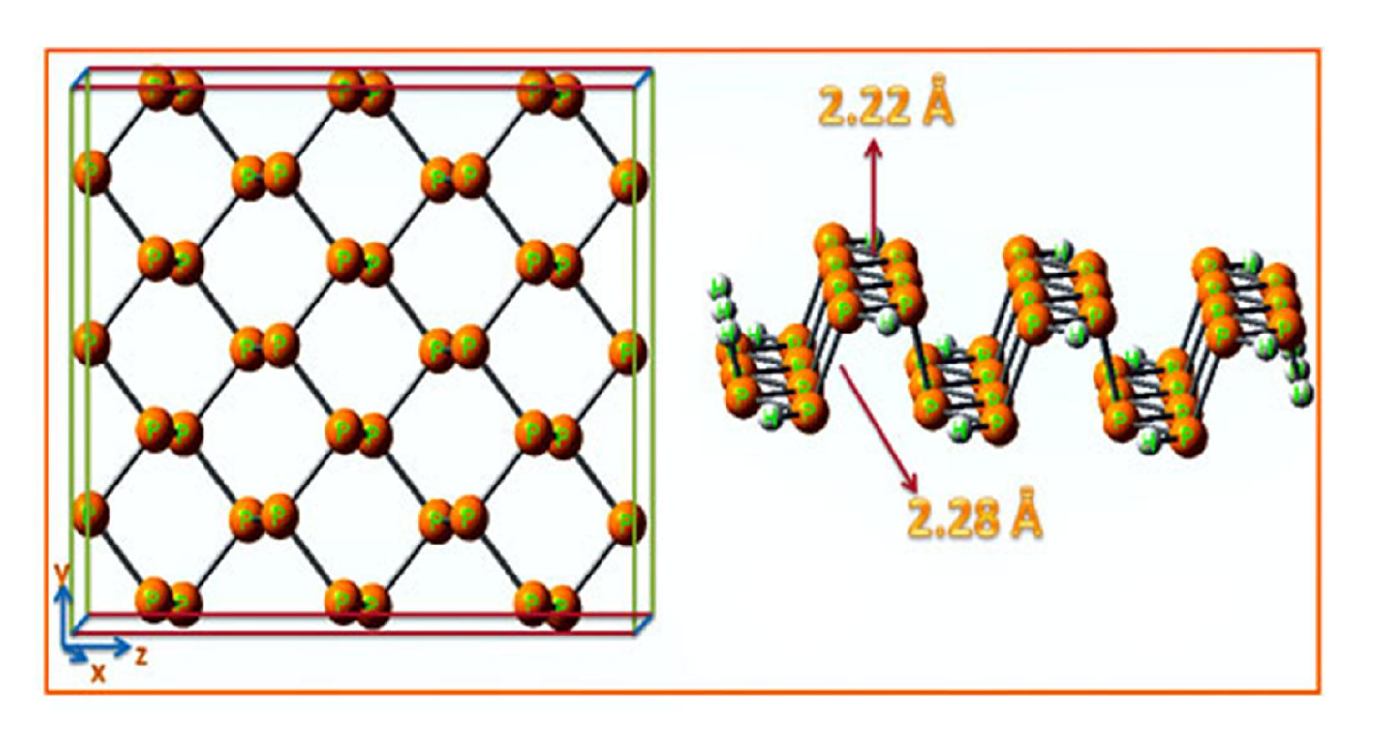}}
\caption{(Colour online) Hydrogenated phosphorene nanosheet.} \label{fig-s1}
\end{figure}

\begin{figure}[!b]
\centerline{\includegraphics[width=0.6\textwidth]{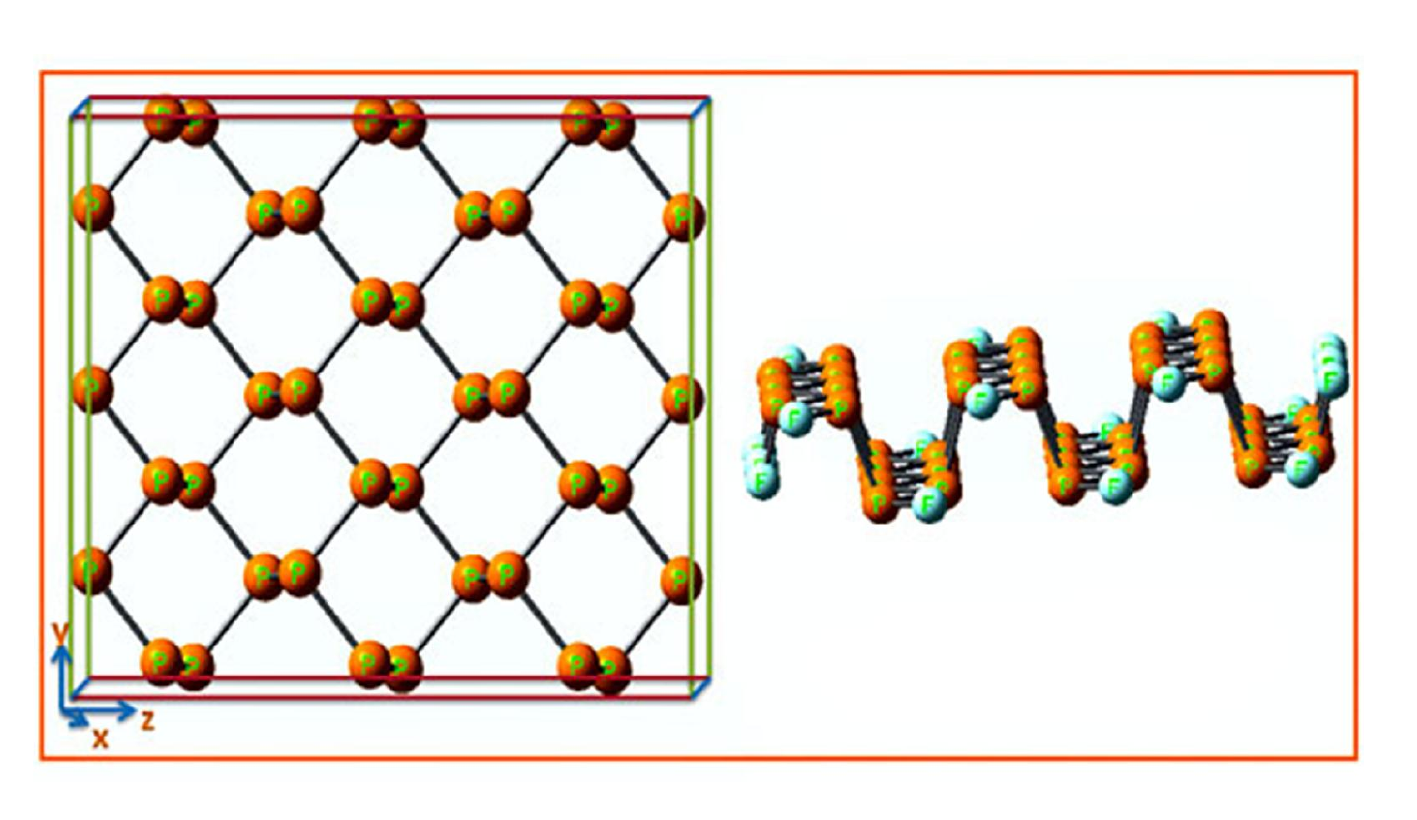}}
\caption{(Colour online) Fluorinated phosphorene nanosheet.} \label{fig-s2}
\end{figure}

The prime motivation of the proposed work is to investigate the formation energy, dipole moment, HOMO-LUMO gap and interaction behaviour of NO$_2$ and NH$_3$ on H-PNS and F-PNS. Figures~\ref{fig-s1} and \ref{fig-s2} refer to H-PNS and F-PNS material, respectively, which is suitable for gas/solid interaction in nanostructures. The lattice constant of black phosphorene is obtained to be $a = 4.41$~{\AA}  and  $b = 3.27$~\AA, which is authenticated with the published work \cite{40}. The H-PNS has forty-two P-atoms, and eighteen H-atoms are passivated on PNS in order to eliminate dangling-bonds. Similarly, eighteen fluorine atoms are passivated with PNS in fluorinated PNS. Furthermore, the termination of PNS with hydrogen and fluorine also influences the interaction behaviour of NO$_2$ and NH$_3$ on PNS material.


\subsection {Electronic properties and geometric stability of phosphorene nanosheets}

The geometric stability of H-PNS and F-PNS was investigated with regard to the formation energy ($E_\text{form}$) as shown in equation (\ref{3.1}) and (\ref{3.2}),
\begin{equation}
	E_\text{form} = \dfrac{1}{n}[E \text{(H-PNS)} - x E\text{(P)} - y E(\text{H})],
	\label{3.1} 
\end{equation}
\begin{equation}
	E_\text{form} = \dfrac{1}{n}[E \text{(F-PNS)} - x E(\text{P}) - y E(\text{F})], 
	\label{3.2} 
\end{equation}
where $E$(H-PNS) and $E$(F-PNS) represents the overall energy of H-PNS and F-PNS material, respectively. $E$(P) refers to the energy of single P atoms, $E$(H) and $E$(F) illustrates the corresponding isolated energy of H and F atom. $n$ denotes the overall atoms including hydrogen or fluorine and phosphorus in PNS base material. Further, $x$ and $y$ refers to the total number of P-atoms and terminated atoms (namely H or F), respectively. The point group (PG), dipole moment (DM) and formation energy of H-PNS and F-PNS are given in table~\ref{table 1}. The $E_\text{form}$ of H-PNS and F-PNS are noticed to be $-3.16$ and $-3.21$~eV, respectively. Clearly, the geometric stability of phosphorene base material slightly improves with the passivation of fluorine atoms. The DP delivers the perception of the dispersal of charges along PNS material. The hydrogenated as well as fluorinated PNS have DM value of zero. It is inferred that the even charge distribution in phosphorene material is perceived on both hydrogenated and fluorinated PNS. C$_\text{2H}$ point group is obtained on both H-PNS and F-PNS base material, which exhibits horizontal reflection plane symmetry.

\begin{table}[!t]
\caption{ Dipole moment (DM), formation energy ($E_\text{form}$), and point group (PG) of phosphorene  nanosheets.}
  \label{table 1}
  \centering
\vspace{2ex}
  \begin{tabular}{|c|c|c|c|}
  \hline\hline
  \raisebox{-1.50ex}[0cm][0cm]{Nanostructures}	&$E_\text{form}$ &	DM & PG \\
     & (eV)&(debye)	 & \\
 
  \hline\hline
 Hydrogenated phosphorene	&$-3.16$	&0	&C$_\text{2H}$ \\
\hline
  Fluorinated phosphorene	&$-3.21$	&0	&C$_\text{2H}$ \\
  \hline\hline
  \end{tabular}
  \end{table}
	\normalsize

\begin{figure}[!b]
\vspace{-3mm}
\centerline{\includegraphics[width=0.65\textwidth]{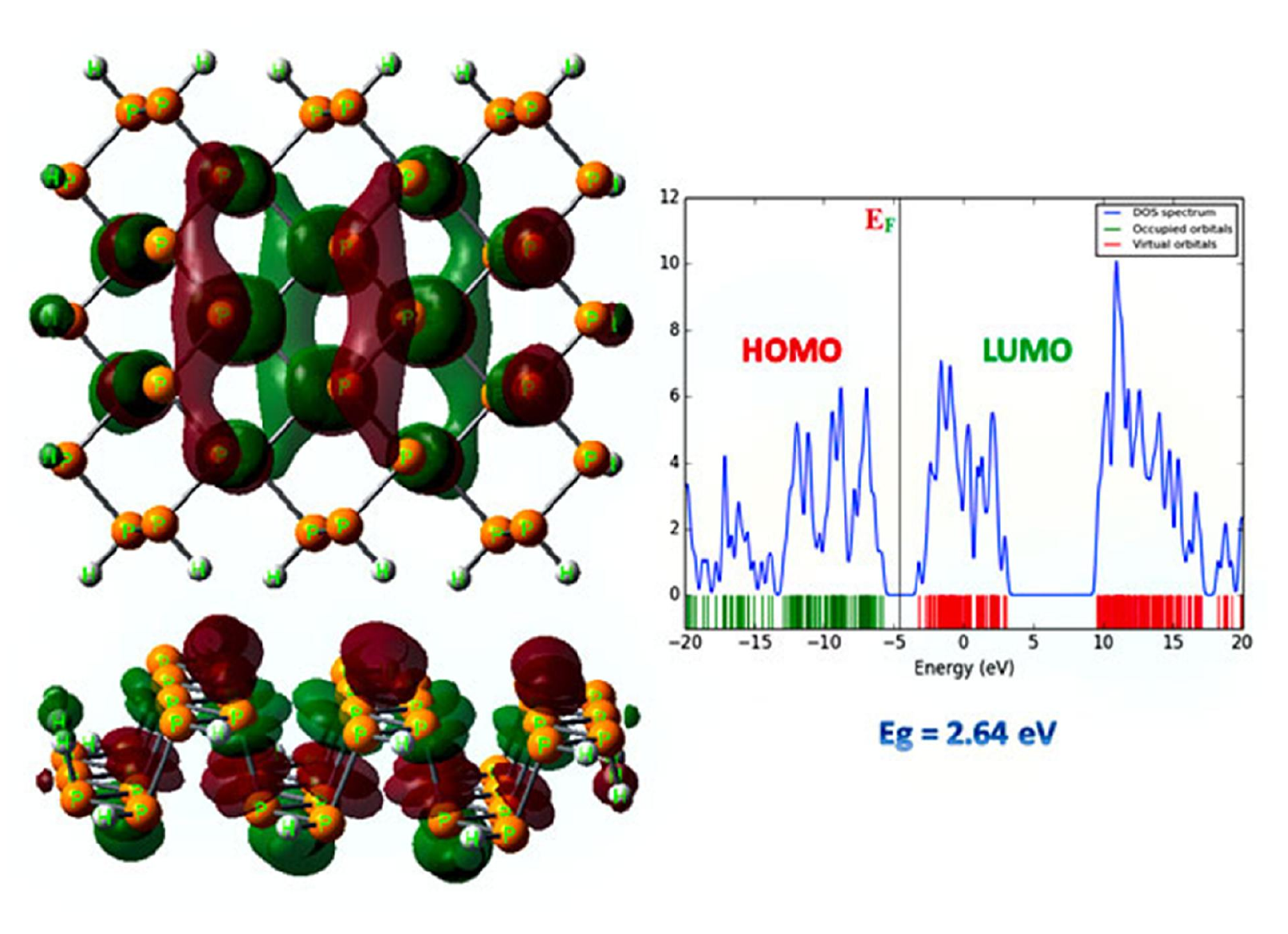}}
\caption{(Colour online) HOMO-LUMO visualization and density of states of hydrogenated PNS.} \label{fig-s3}
\end{figure}
\begin{figure}[!t]
\centerline{\includegraphics[width=0.65\textwidth]{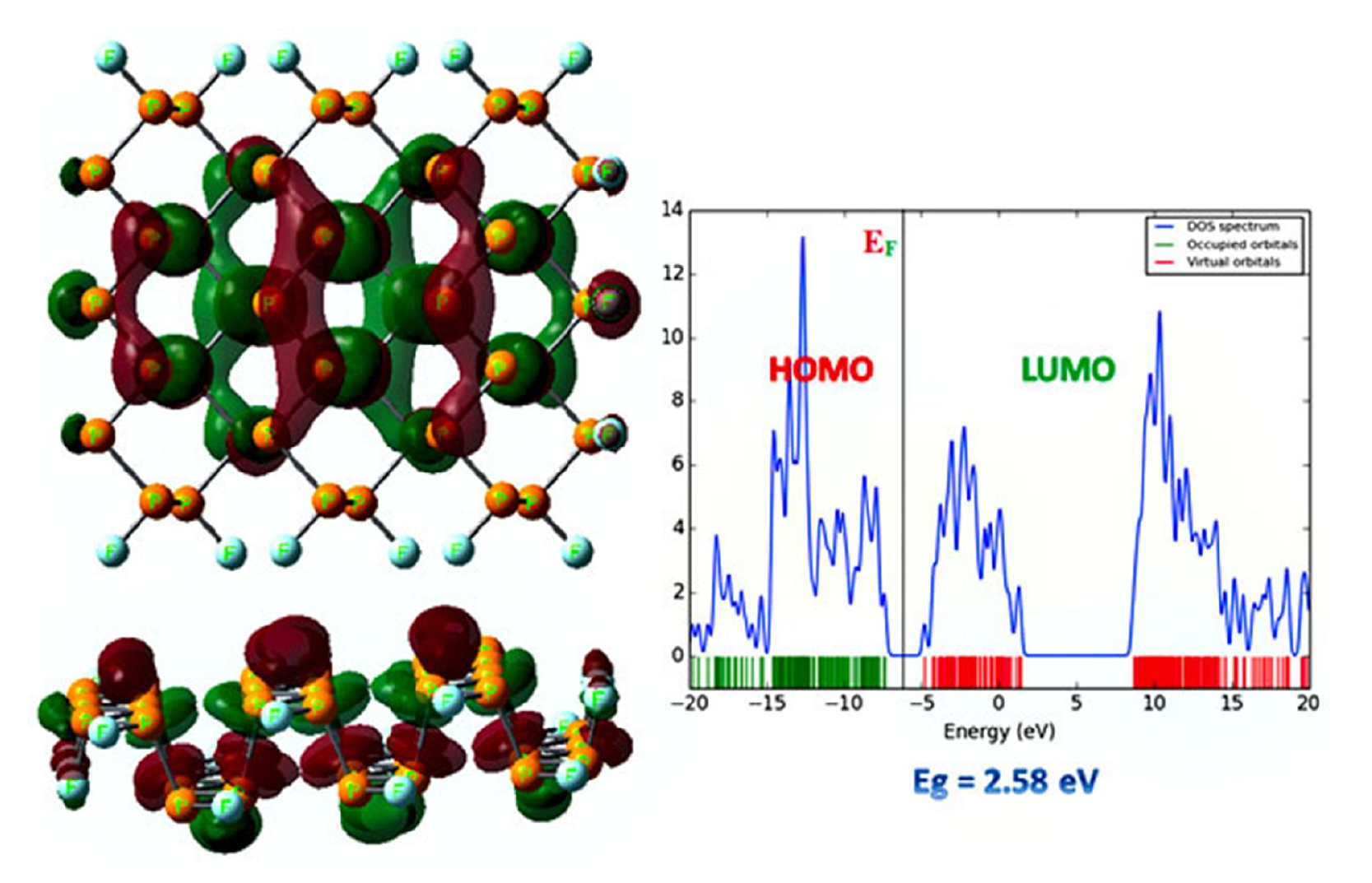}}
\caption{(Colour online) HOMO-LUMO visualization and density of states of fluorinated PNS.} \label{fig-s4}
\end{figure}

The electronic characteristics of H-PNS and F-PNS are inspected in terms of LUMO and HOMO levels \cite{41,42}. The HOMO-LUMO gap of H-PNS and F-PNS are found to be 2.64 and 2.58 eV, respectively. Boukhvalov et al. \cite{43} reported the stability and chemical modification of phosphorene with the influence of fluorine and hydrogen. They calculated the energy gap value for two different structural configurations of fully hydrogenated PNS, which is noticed to be 2.29 eV and 2.64 eV, respectively. Besides, for fluorinated PNS, the reported band gap values are 2.27 and 1.82 eV. The experimental band gap value for bulk black phosphorus is 0.35 eV, whilst for single layer phosphorene, it is found to be around 2.0 eV \cite{44,45,46,47}. Moreover, the band gap of PNS in the present work is also observed in this range, which is consistent with the previously reported experimental and theoretical work. In addition, the HOMO-LUMO gap variation arose owing to the orbital coincidence of H and F with P atoms in PNS material. The localized electronic states of PNS influence the DOS spectrum in various energy intervals through PNS. The pictorial representation of the HOMO-LUMO gap with HOMO and LUMO levels and DOS spectrum of hydrogenated and fluorinated PNS are shown in figure~\ref{fig-s3} and \ref{fig-s4}. In all the cases, peak maxima ($P_\text{max}$) are observed to be high in LUMO level, which is the most required condition for the interaction of hazardous gases. This is because in this condition, the electrons can freely transit between NO$_2$ or NH$_3$ gas molecules and PNS material. The $P_\text{max}$ appears owing to the orbital intersection between P-atoms and H or F-atoms in PNS along different energy intervals.

\subsection {Interaction behaviour of NO$_2$ and NH$_3$ on PNS}
\begin{figure}[!b]
\centerline{\includegraphics[width=0.65\textwidth]{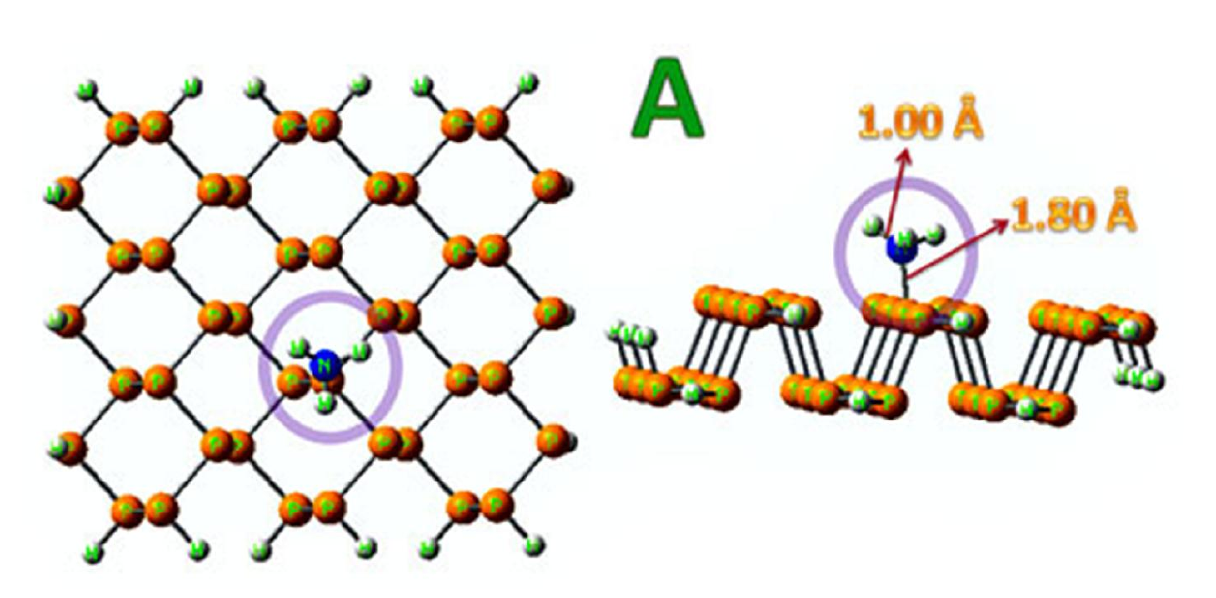}}
\caption{(Colour online) NH$_{3}$ interacted on H-PNS --- orientation `A'.} \label{fig-s5}
\end{figure}

\begin{figure}[!t]
\centerline{\includegraphics[width=0.65\textwidth]{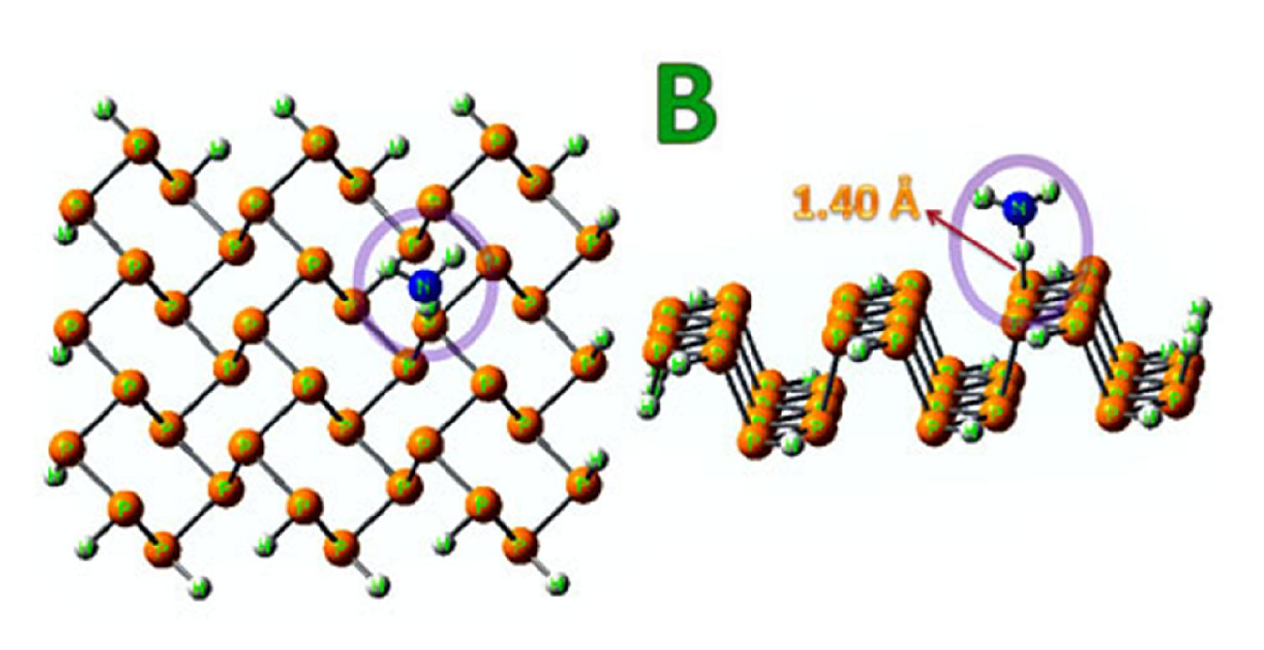}}
\caption{(Colour online) NH$_{3}$ interacted on H-PNS --- orientation `B'.} \label{fig-s6}
\end{figure}

\begin{figure}[!t]
\centerline{\includegraphics[width=0.65\textwidth]{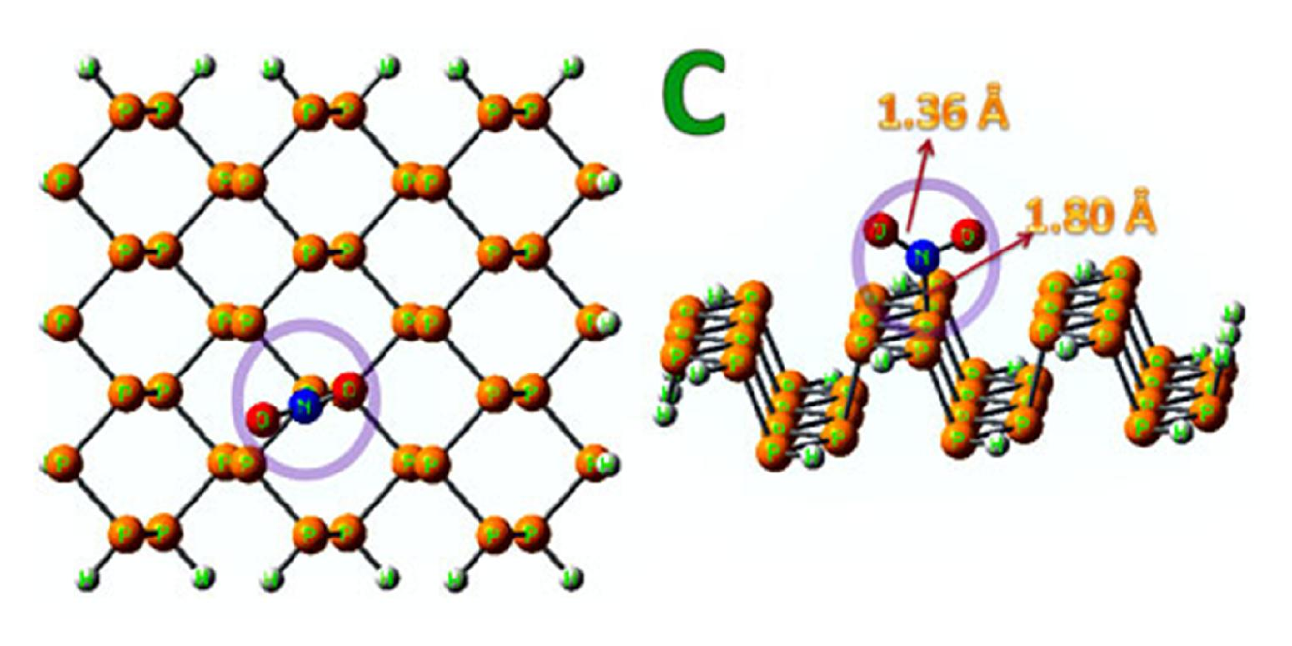}}
\caption{(Colour online) NO$_2$ interacted on H-PNS --- orientation `C'.} \label{fig-s7}
\end{figure}

\begin{figure}[!t]
\centerline{\includegraphics[width=0.65\textwidth]{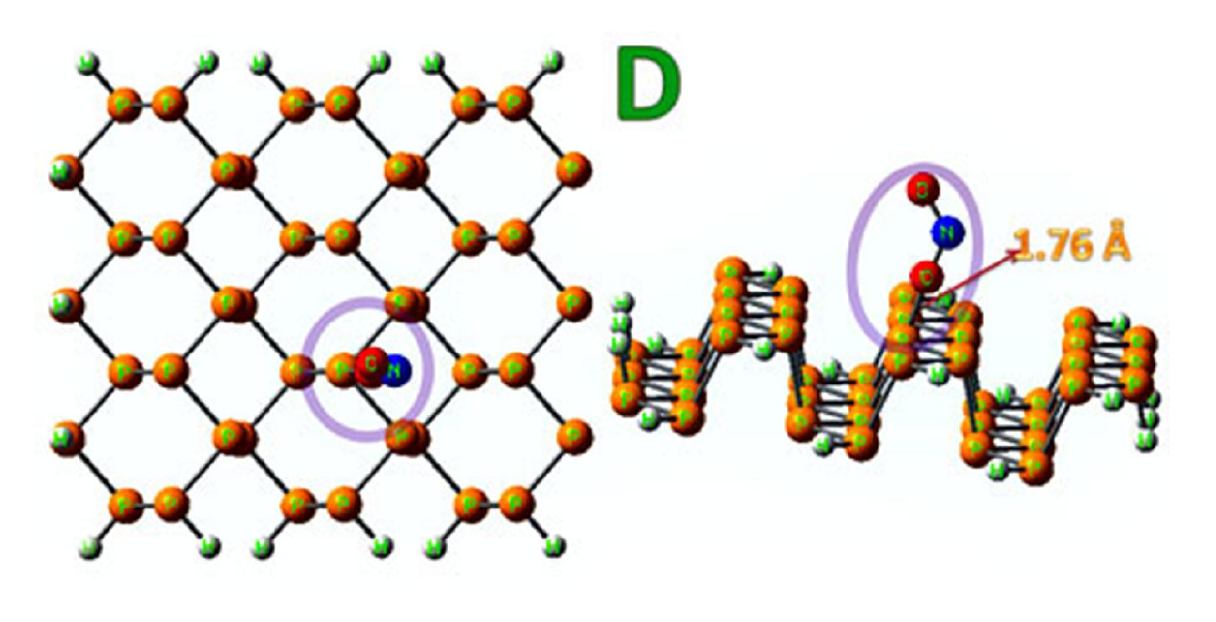}}
\caption{(Colour online) NO$_2$ interacted on H-PNS --- orientation `D'.} \label{fig-s8}
\end{figure}

\begin{figure}[!t]
\centerline{\includegraphics[width=0.65\textwidth]{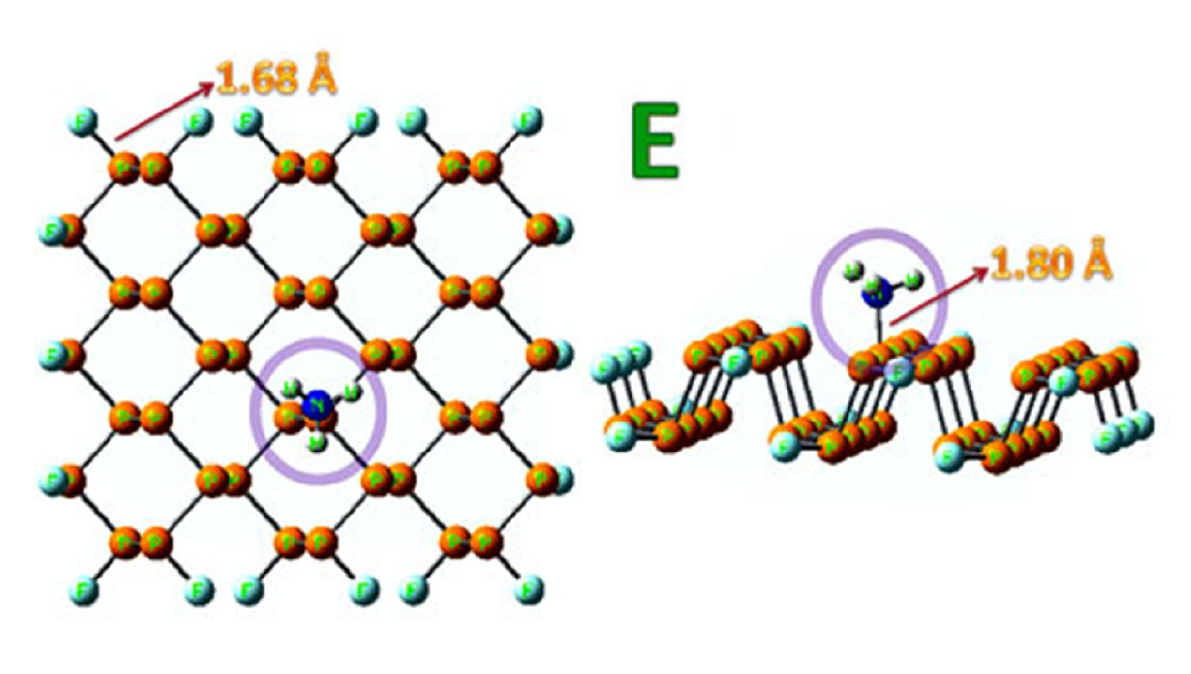}}
\caption{(Colour online) NH$_3$ interacted on F-PNS --- orientation `E'.} \label{fig-s9}
\end{figure}

\begin{figure}[!t]
\centerline{\includegraphics[width=0.65\textwidth]{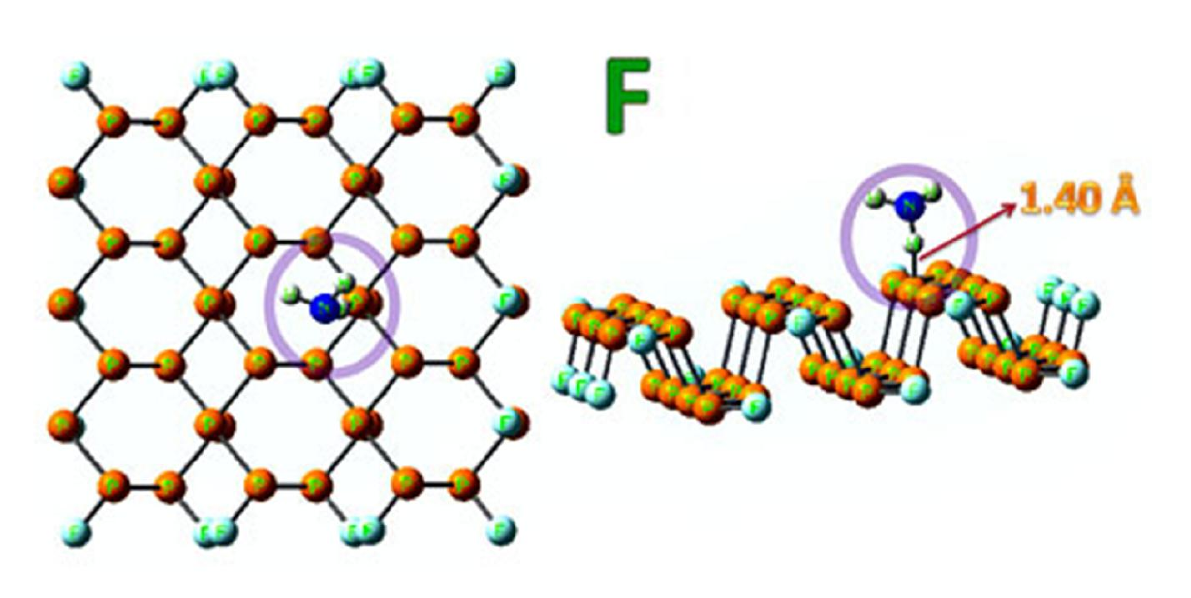}}
\caption{(Colour online) NH$_3$ interacted on F-PNS --- orientation `F'.} \label{fig-s10}
\end{figure}

\begin{figure}[!t]
\centerline{\includegraphics[width=0.65\textwidth]{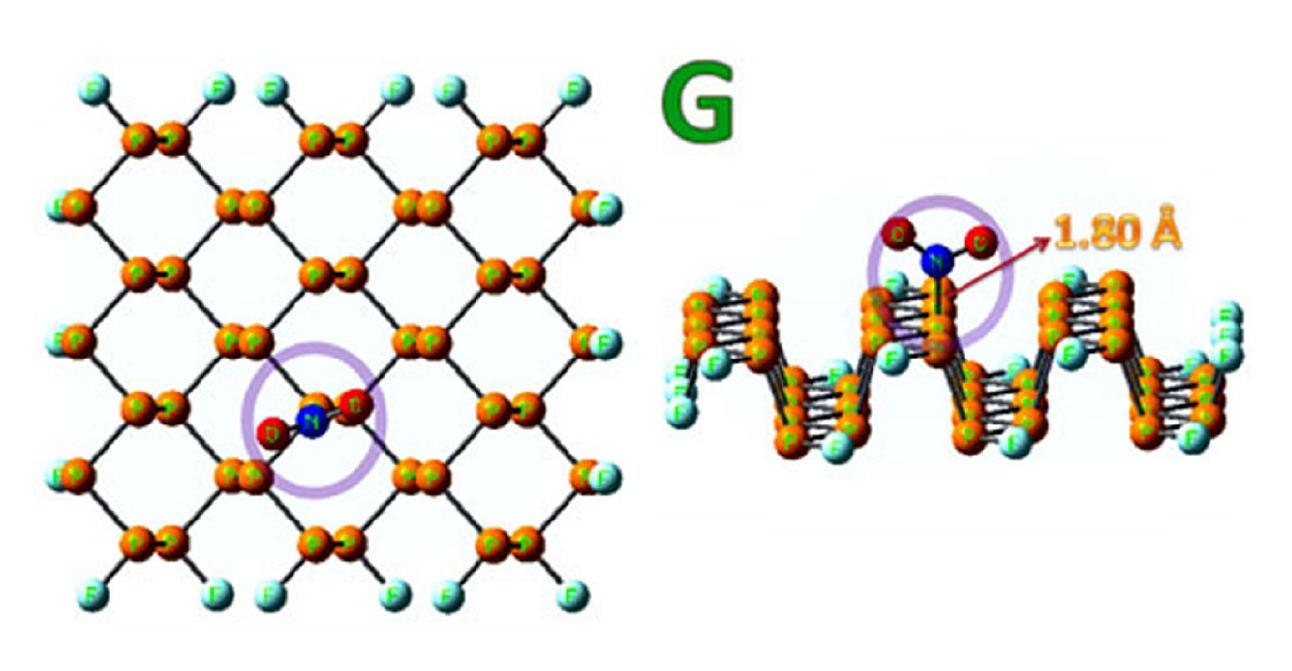}}
\caption{(Colour online) NO$_2$ interacted on F-PNS --- orientation `G'.} \label{fig-s11}
\end{figure}

\begin{figure}[!t]
\centerline{\includegraphics[width=0.65\textwidth]{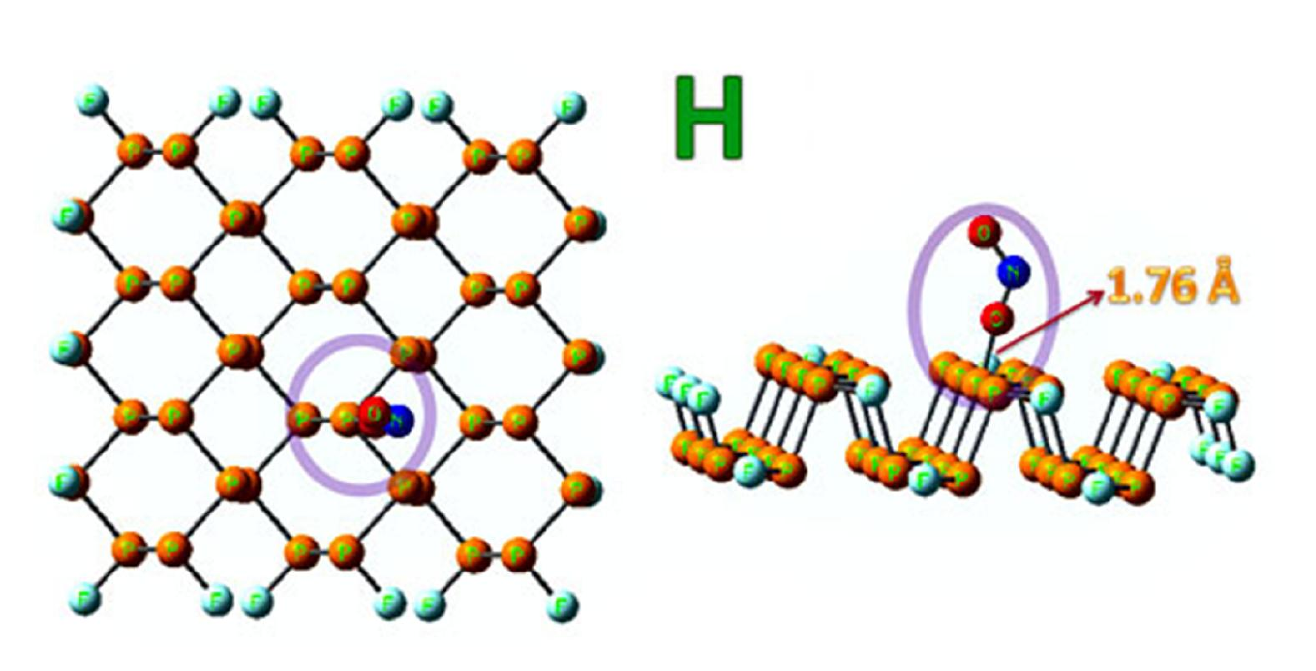}}
\caption{(Colour online) NO$_2$ interacted on F-PNS --- orientation `H'.} \label{fig-s12}
\end{figure}

At the beginning stage, to study the NO$_2$ and NH$_3$ interaction behaviour on phosphorene base material, NO$_2$, and NH$_3$ molecules should be studied in the gas-phase. The optimized bond distance between the P-atoms in PNS is found to be 2.22 \AA. The calculated bond length between the N-H in NH$_3$ and N-O in NO$_2$ is $1.00$~{\AA} and 1.36~\AA, respectively. Figures~\ref{fig-s5}, \ref{fig-s6} and \ref{fig-s7}, \ref{fig-s8} represent the adsorption of NH$_3$ and NO$_2$ gases on various sites of H-PNS, respectively. In that order, figures~\ref{fig-s9}, \ref{fig-s10} and \ref{fig-s11}, \ref{fig-s12} refer to the adsorption of NH$_3$ and NO$_2$ gases on different sites of fluorinated PNS.


The interaction of nitrogen and hydrogen atom in NH$_3$ molecules interacted on phosphorus atom in H-PNS is referred to as orientations A and B, respectively. Orientations C and D represent the interaction of N and O-atom in NO$_2$ gas interacting on phosphorus atom in hydrogenated PNS, respectively. Similarly, orientations E--H refer to the interaction of NH$_3$ and NO$_2$ molecules adsorbed on phosphorus atom in fluorinated PNS. 

\begin{table}[!b]
  \caption{ NBO charge transfer, HOMO-LUMO gap, adsorption energy and average HOMO-LUMO gap changes of phosphorene nanosheets.}
  \label{table 2}
  \centering
  \vspace{2ex}
  \scriptsize\begin{tabular}{|c|c|c|c|c|c|c|c|c|}
  \hline\hline
  Phosphorene nanostructures  	& \raisebox{-1.50ex}[0cm][0cm]{DFT $E_\text{ad}$ (eV)} 	& \raisebox{-1.50ex}[0cm][0cm]{D-DFT $E_\text{ad}$ (eV)}  & \raisebox{-1.50ex}[0cm][0cm]{Q ($e$)} 	& \raisebox{-1.50ex}[0cm][0cm]{$E_\text{HOMO}$} 	& \raisebox{-1.50ex}[0cm][0cm]{$E_\text{FL}$ (eV)} 	&\raisebox{-1.50ex}[0cm][0cm]{$E_\text{LUMO}$} 	&\raisebox{-1.50ex}[0cm][0cm]{$E_\text{g}$ (eV)} 	&\raisebox{-1.50ex}[0cm][0cm]{$E_\text{g}^\text{a} \ \%$}  \\
 and orientation &&&&&&&&\\
	\hline\hline
Hydrogenated phosphorene 	&$-$	&$-$	&$-$ &$-5.83$	&$-4.51$	&$-3.19$	&2.64	&$-$	\\
nanosheet&&&&&&&&\\
\hline
A	&$-2.18$	& $-2.213$ &0.405	&$-4.43$	&$-3.73$	&$-3.03$	&1.40	&46.97 \\

\hline
B	&$-2.18$	& $-2.207$ &0.140	&$-5.66$	&$-4.38$	&$-3.10$	&2.56	&3.03 \\
\hline
C	&$-1.36$	&$-1.401$ &$-0.322$	&$-5.38$	&$-4.27$	&$-3.16$	&2.22	&15.91 \\
\hline
D	&$-1.90$	&$-1.989$ &$-0.359$	&$-5.56$	&$-8.195$	&$-7.26$	&2.46	&6.82 \\
\hline
Fluorinated phosphorene 	&$-$	&$-$	&$-$ &$-7.37$	&$-6.08$	&$-4.79$	&2.58	&$-$ \\
nanosheet&&&&&&&&\\
\hline
E	&$-2.18$	&$-2.222$ &0.427	&$-5.94$	&$-5.28$	&$-4.62$	&1.32	&48.84 \\
\hline
F	&$-2.45$	&$-2.500$ &0.172	&$-7.14$	&$-5.915$	&$-4.69$	&2.45	&5.04 \\
\hline
G	&$-1.63$	&$-1.658$ &$-0.295$	&$-6.89$	&$-5.79$	&$-4.69$	&2.20	&14.73 \\
\hline
H	&$-2.18$	&$-2.197$ &$-0.302$	&$-7.02$	&$-5.845$	&$-4.67$	&2.35	&8.91 \\
\hline\hline
  \end{tabular}
  \end{table}
\normalsize

The corresponding adsorption energy ($E_\text{ad}$) H-PNS and F-PNS upon the interaction of NH$_3$ and NO$_2$ gases can be determined using equations~(\ref{3.3}) and (\ref{3.4})
\begin{equation}
E_\text{ad} = E(\text{PNS/NH}_3) - E(\text{PNS}) - E(\text{NH}_3) + E(\text{BSSE}),
\label{3.3}	
\end{equation}
\begin{equation}
E_\text{ad} = E(\text{PNS/NO}_2) - E(\text{PNS}) - E(\text{NO}_2) + E(\text{BSSE}),
\label{3.4}	
\end{equation}
where $E$(PNS/NH$_3$) and $E$(PNS/NO$_2$) represent the energy of PNS/NH$_3$ and PNS/NO$_2$ complex, respectively. The isolated energies of PNS are stated as $E$(PNS). $E$(NH$_3$) and $E$(NO$_2$) signify the respective isolated energy of NH$_3$ and NO$_2$ toxic gas molecules. The basis-set-superposition-error (BSSE) \cite{48} has been calculated by the counterpoise method to neglect the overlap effects on the basis sets. When NH$_3$ or NO$_2$ target gas molecules interacted on PNS material, the negative scale of $E_\text{ad}$ was obtained which infers a strong interaction of NH$_3$/NO$_2$ on PNS. Importantly, the aforementioned adsorption site from A to H shows the negative scale of adsorption energy, which confirms that the NO$_2$ and NH$_3$ molecules strongly interact on PNS. The calculated $E_\text{ad}$ of hydrogenated PNS for orientations A--D is noticed to be $-2.18$, $-2.18$, $-1.36$ and $-1.9$~eV, respectively. The $E_\text{ad}$ of F-PNS material for the interaction sites E--H is found to be $-2.18$, $-2.45$, $-1.63$ and $-2.18$~eV, respectively. Moreover, only a small variation is noticed in $E_\text{ad}$ for all aforementioned interactions sites with the inclusion of Grimme dispersion correction (D-DFT). Furthermore, the energy gap of PNS material contracts owing to the interaction of NO$_2$ and NH$_3$ molecules on both H-PNS and F-PNS due to the influence of electronic configuration of these two gas molecules on passivated hydrogen and fluorine PNS \cite{49}. Thus, the conductivity of PNS increases. The variation in conductivity of H-PNS and F-PNS confirms that the electrons are relocated between the PNS material and toxic gas molecules giving rise to an enhanced conductivity. Using a simple two-probe method, the changes in the current can be analyzed across PNS, and it is directly related to the amount of NO$_2$ and NH$_3$ gas molecules in the air. In this work, the deviation in the HOMO-LUMO gap of PNS for the interaction sites A to H is noticed to be 1.4, 2.56, 2.22, 2.46, 1.32, 2.45, 2.2 and 2.35 eV, respectively. Further, the HOMO-LUMO gap variation and adsorption energy reveal that PNS can be utilized for the probing of NO$_2$ and NH$_3$ gas molecules. Cai group \cite{50} studied the adsorption characteristics of a variety of gas molecules. The authors report that the gas molecules are physisorbed on PNS. Kou et al. \cite{25} also confirmed the adsorption properties of different gases on a single layer phosphorene. The authors further concluded that the single layer phosphorene is a promising material for probing toxic gas molecules. From the literature survey, we observed that very limited works were reported on hydrogenated and fluorinated PNS as NO$_2$ and NH$_3$ gas sensors. Moreover, the most suitable interaction site of NO$_2$ and NH$_3$ gas molecules on PNS material can be observed only after studying the percentage of average HOMO-LUMO gap changes related to its pristine PNS. Table~\ref{table 2} represents the HOMO-LUMO gap, percentage of average HOMO-LUMO gap changes, NBO charge transfer and adsorption energy. Furthermore, it is revealed that the most suitable interaction sites for NH$_3$ gases on PNS are A and E. Nevertheless, the most suitable interaction sites for NO$_2$ gas on PNS material are C and G. The interaction of the nitrogen element in both NO$_2$ and NH$_3$ gas molecules, when they get adsorbed on H-PNS and F-PNS, is found to be more favourable on interaction sites. Later, the average HOMO-LUMO gap changes are found to be relatively higher than the other interaction sites. Moreover, phosphorene base material is observed to have a good response and to be sensitive to N-based hazardous gas molecules \cite{25}, which is also validated with the proposed work.

\begin{figure}[!b]
\centerline{\includegraphics[width=0.65\textwidth]{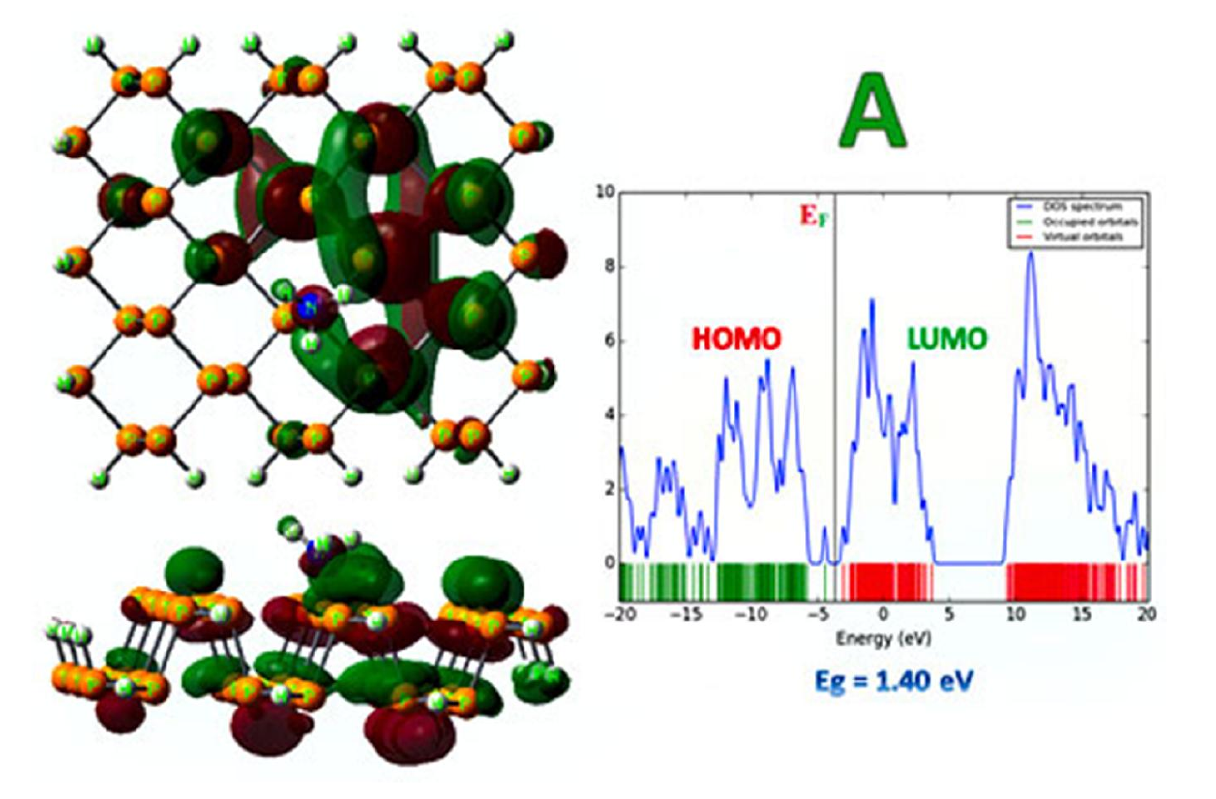}}
\caption{(Colour online) HOMO-LUMO conception and density of states spectrum orientation A.} \label{fig-s13}
\end{figure}

\begin{figure}[!b]
\centerline{\includegraphics[width=0.65\textwidth]{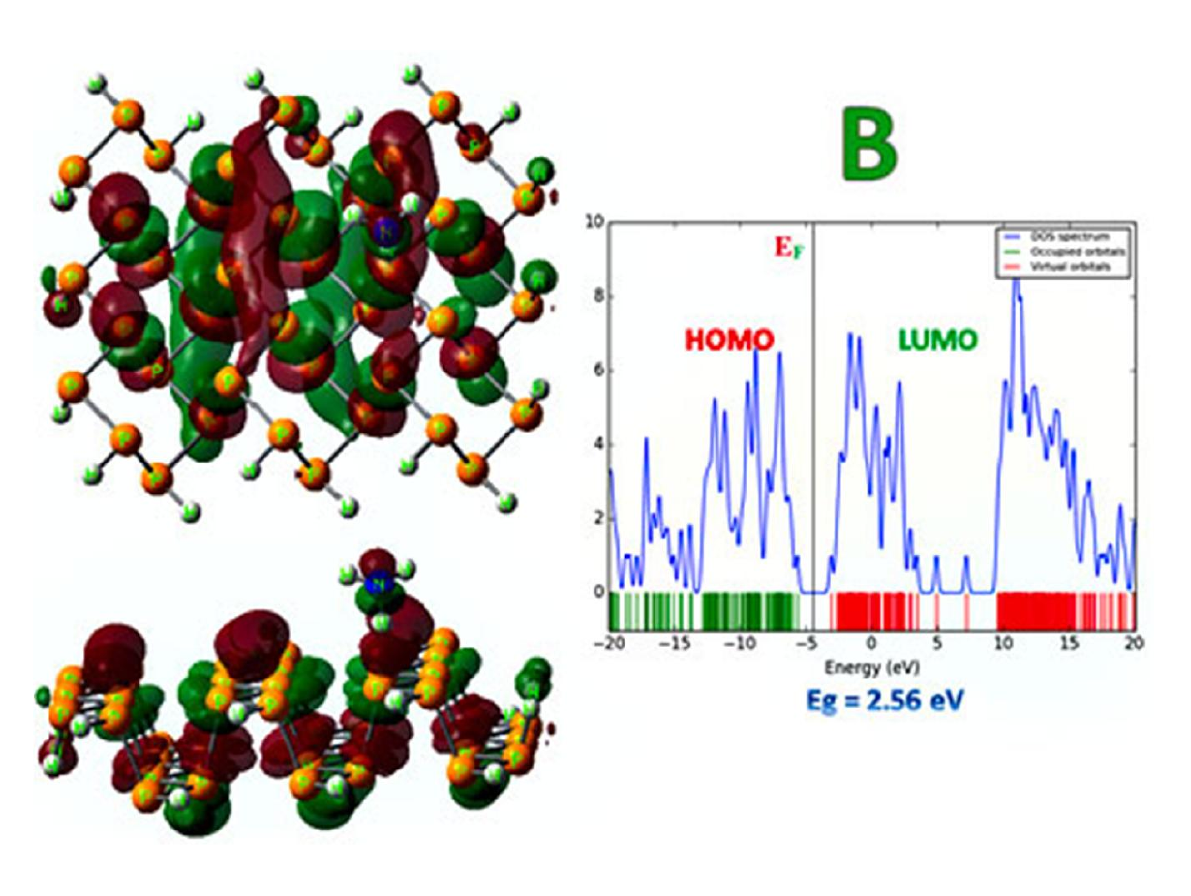}}
\caption{(Colour online) HOMO-LUMO conception and density of states spectrum orientation B.} \label{fig-s14}
\end{figure}

\begin{figure}[!b]
\centerline{\includegraphics[width=0.65\textwidth]{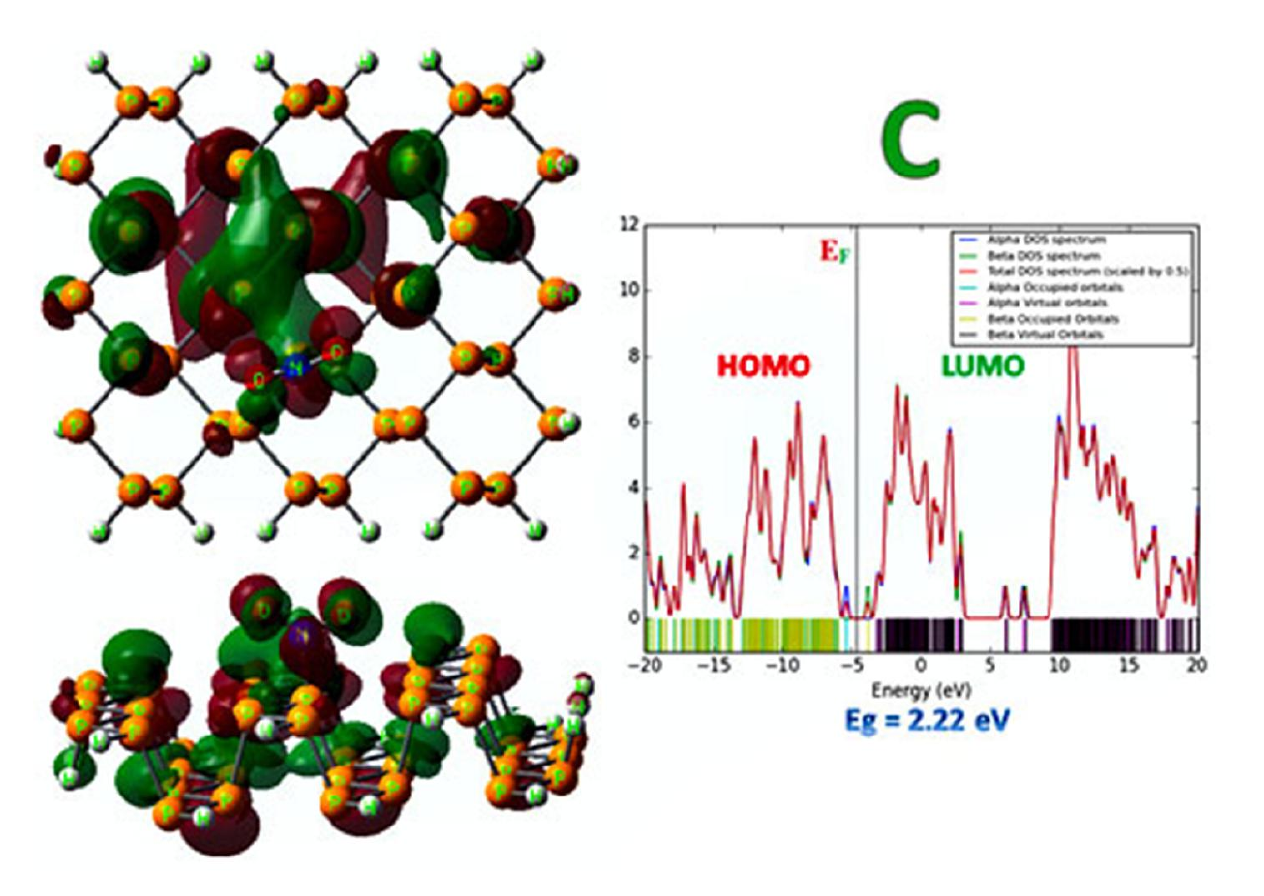}}
\caption{(Colour online) HOMO-LUMO conception and density of states spectrum orientation C.} \label{fig-s15}
\end{figure}

\begin{figure}[!b]
\centerline{\includegraphics[width=0.65\textwidth]{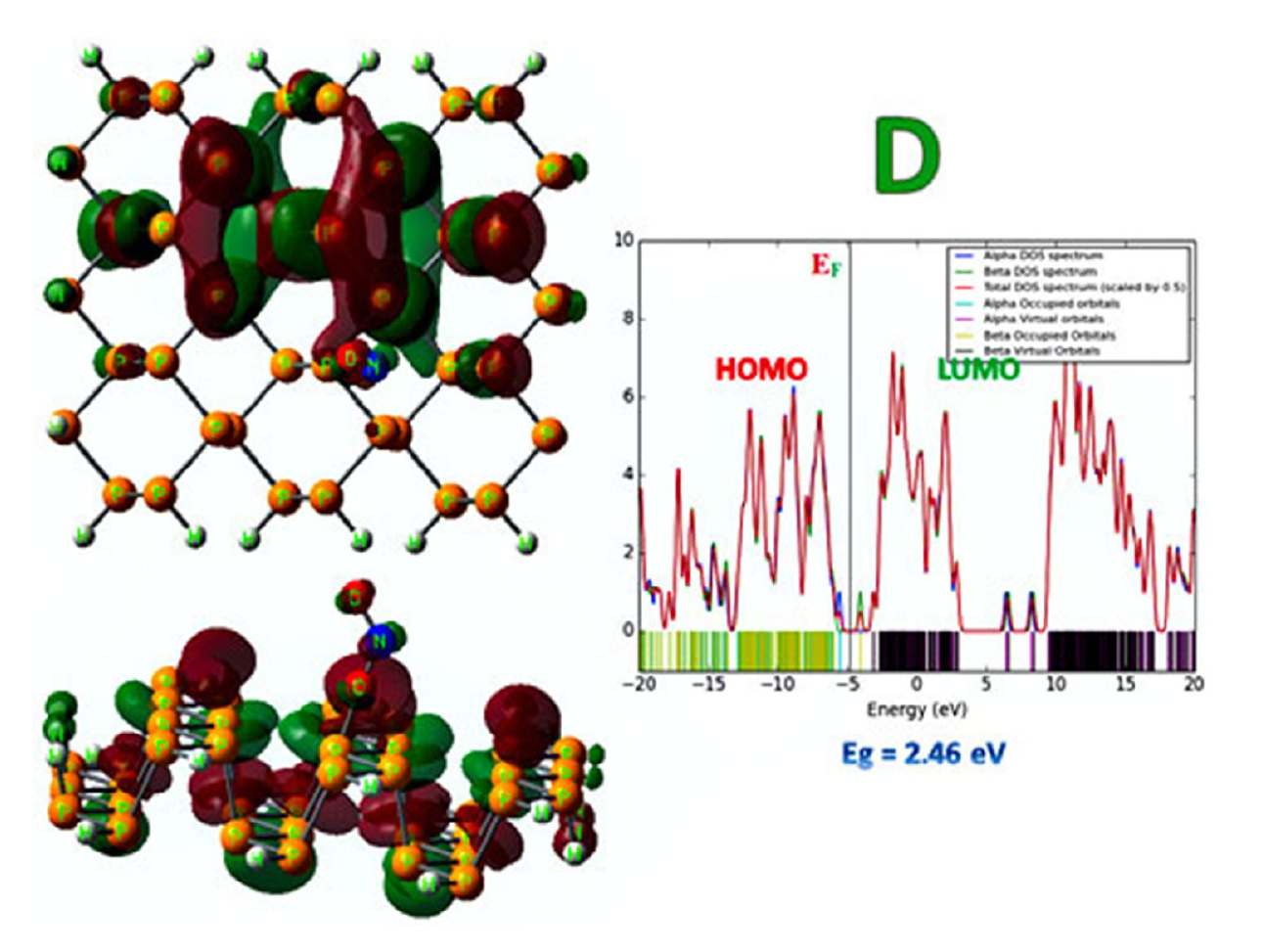}}
\caption{(Colour online) HOMO-LUMO conception and density of states spectrum orientation D.} \label{fig-s16}
\end{figure}

\begin{figure}[!t]
\centerline{\includegraphics[width=0.65\textwidth]{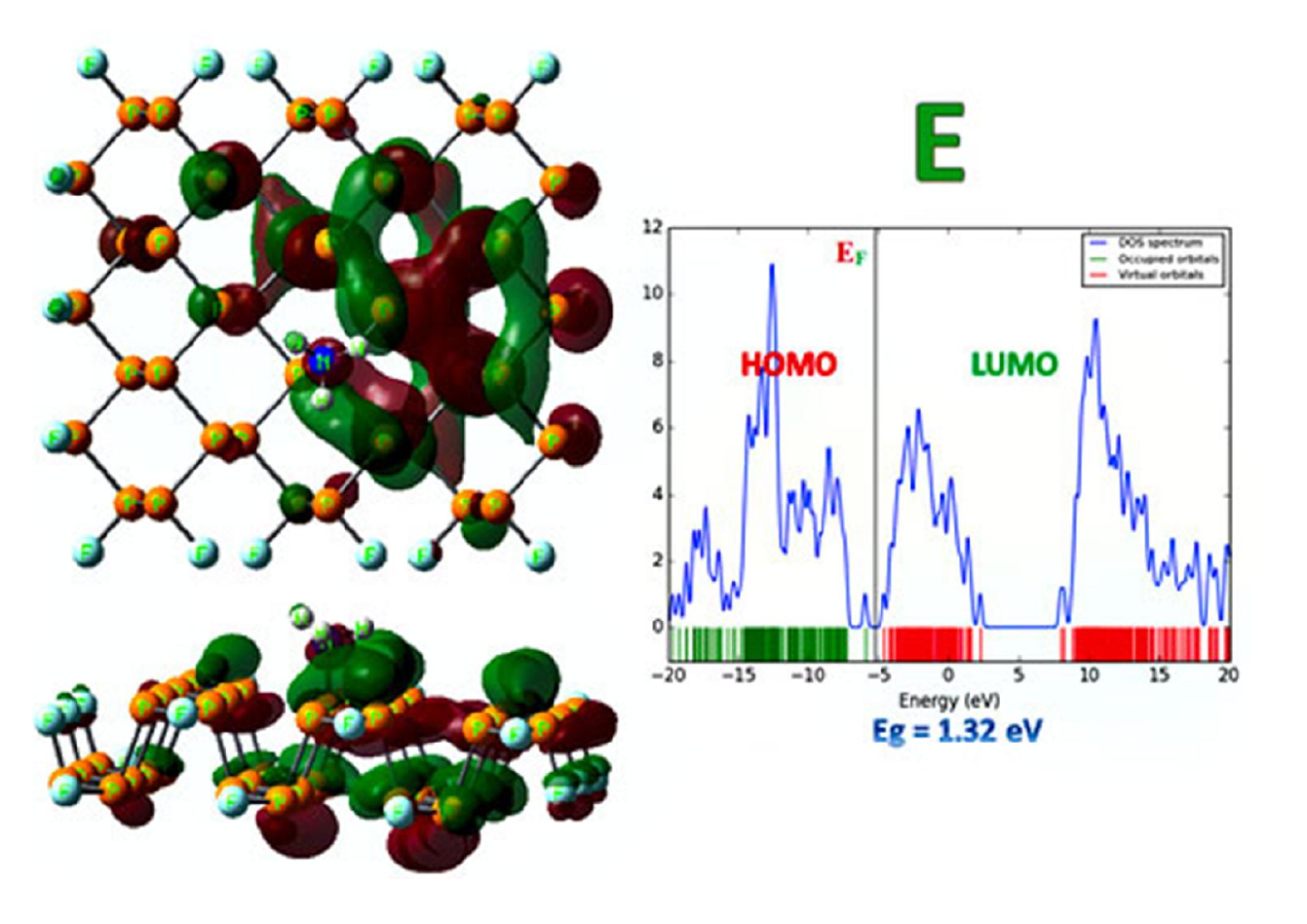}}
\caption{(Colour online) HOMO-LUMO conception and density of states spectrum orientation E.} \label{fig-s17}
\end{figure}

\begin{figure}[!t]
\centerline{\includegraphics[width=0.65\textwidth]{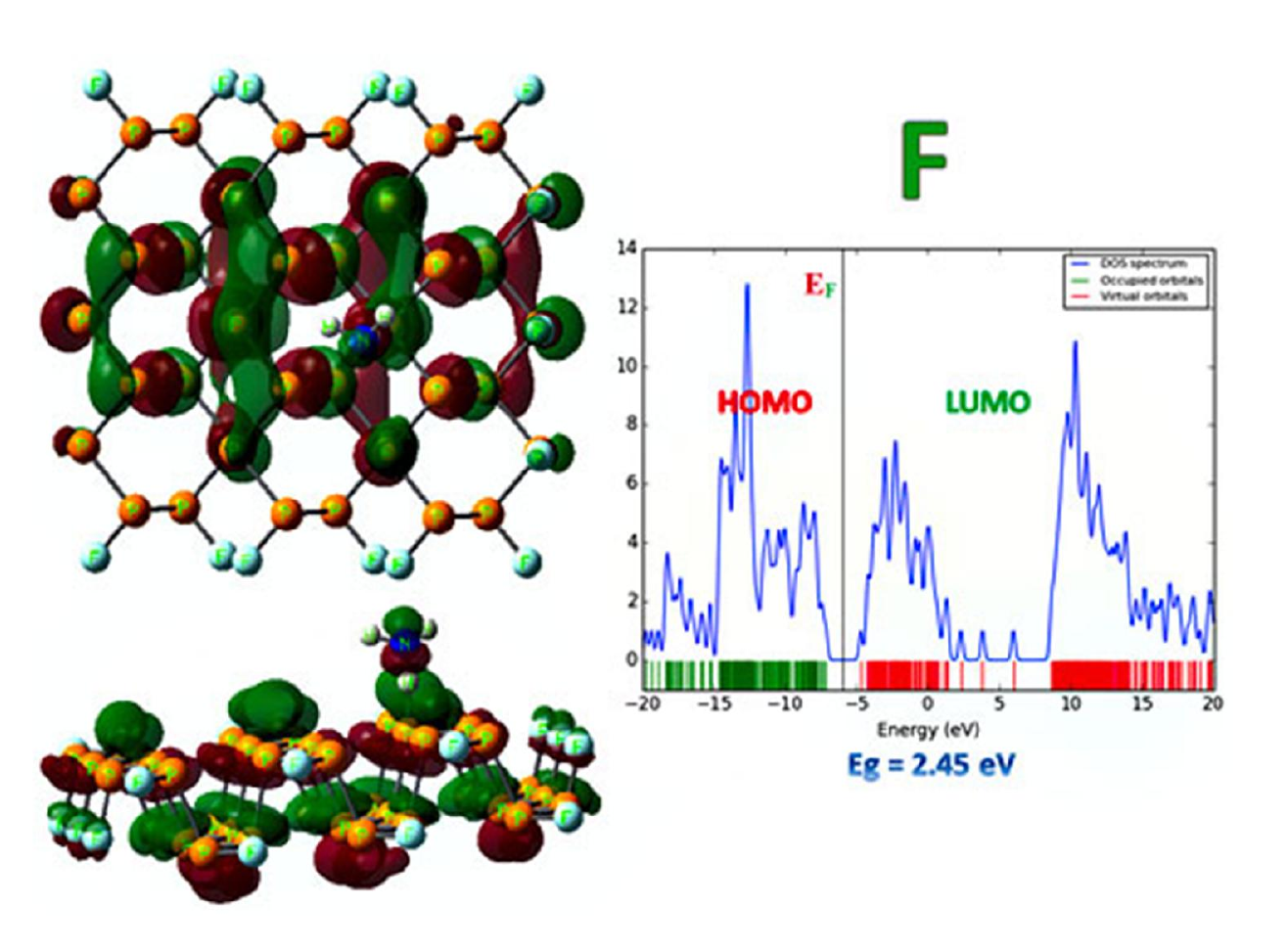}}
\caption{(Colour online) HOMO-LUMO conception and density of states spectrum orientation F.} \label{fig-s18}
\end{figure}

\begin{figure}[!t]
\centerline{\includegraphics[width=0.65\textwidth]{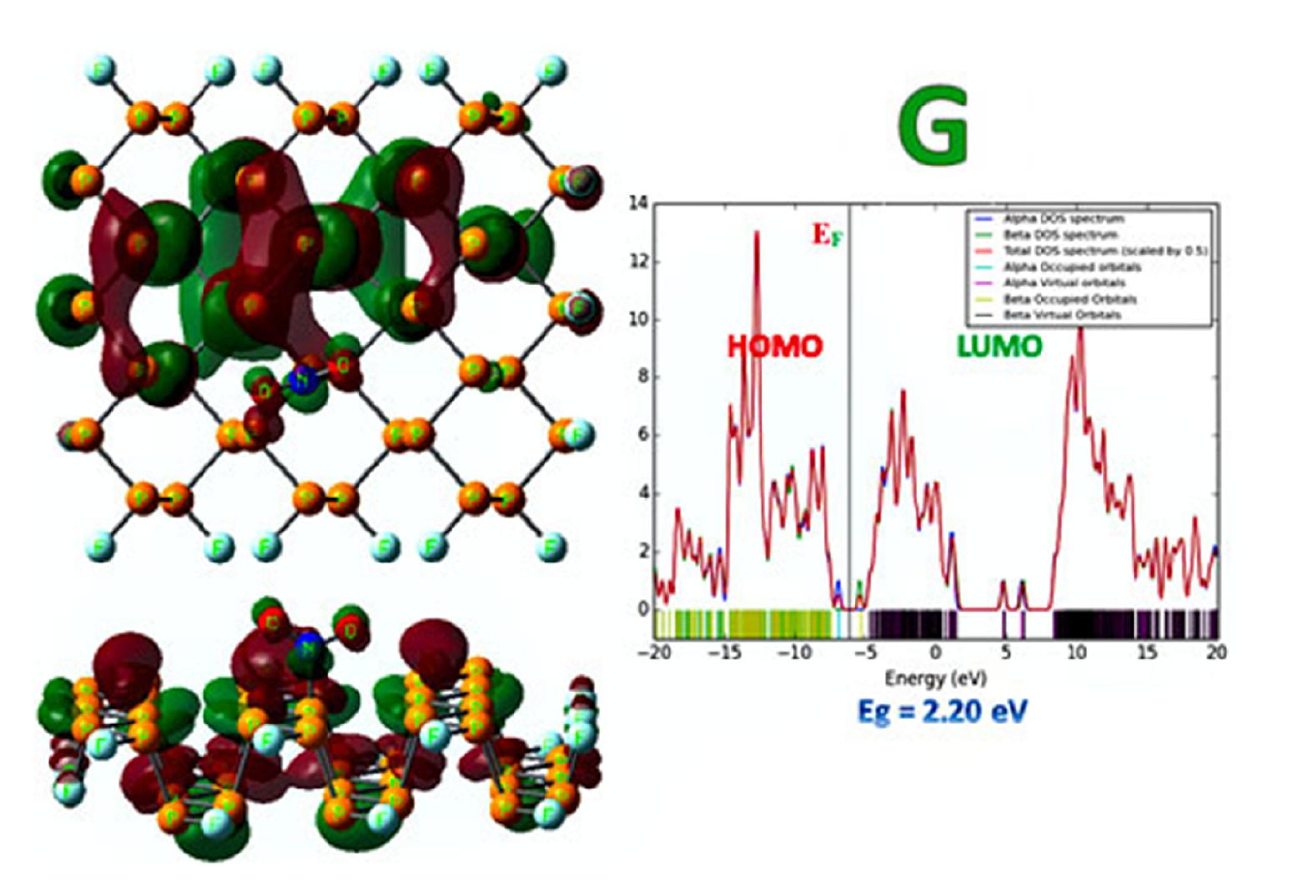}}
\caption{(Colour online) HOMO-LUMO conception and density of states spectrum orientation G.} \label{fig-s19}
\end{figure}

\begin{figure}[!t]
\centerline{\includegraphics[width=0.65\textwidth]{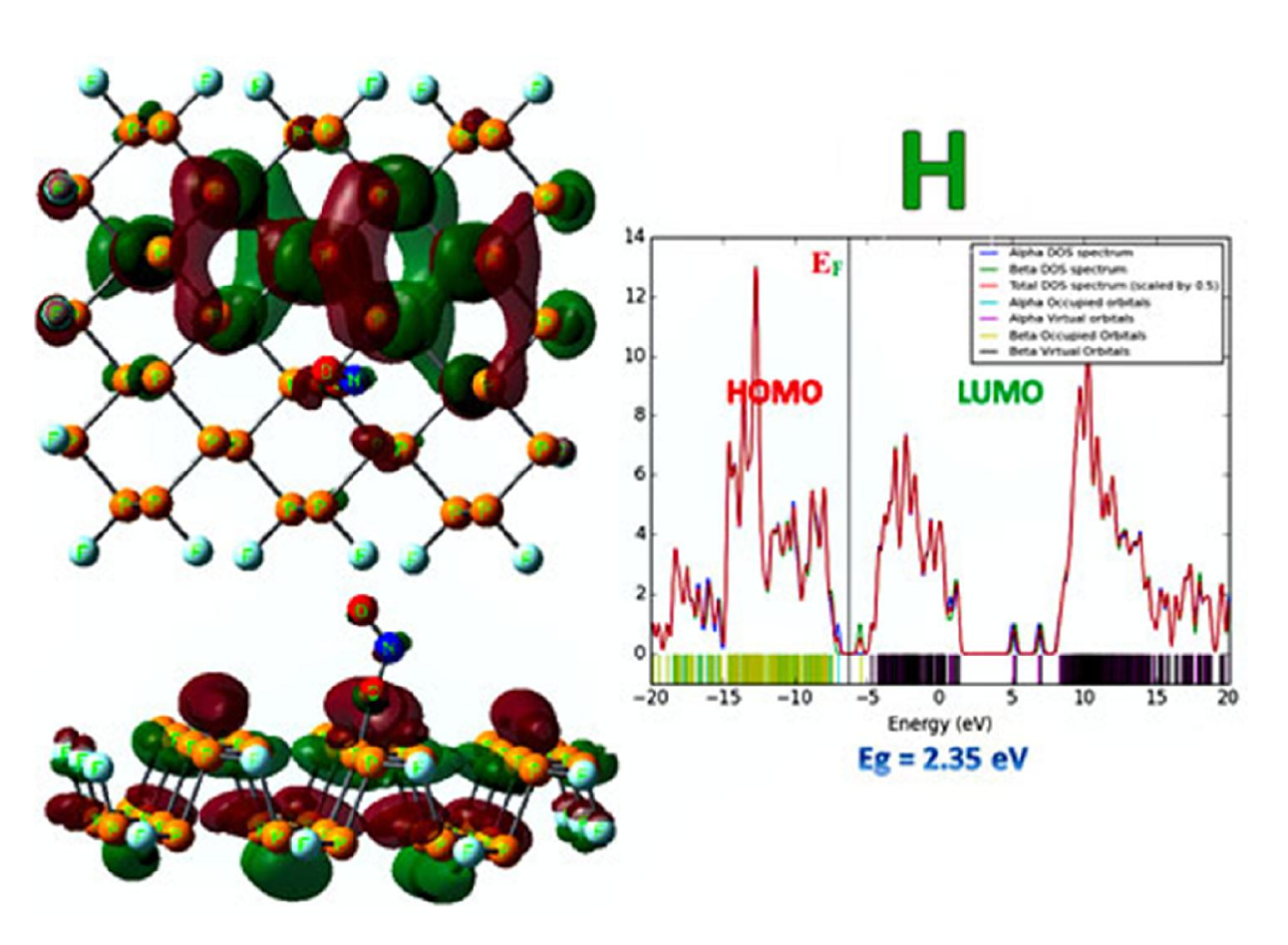}}
\caption{(Colour online) HOMO-LUMO conception and density of states spectrum orientation H.} \label{fig-s20}
\end{figure}

The transfer of electrons between NH$_3$ or NO$_2$ gas molecules and PNS can be described in terms of NBO analysis (Q) \cite{51,52,53}. As we know, the positive magnitude of NBO charge transfer represents the electrons that get transmitted from target NH$_3$/NO$_2$ molecules to PNS; whereas the negative magnitude of NBO charge refers the electrons moving from PNS to NH$_3$/NO$_2$ gases \cite{54,55,56}. The obtained NBO charge values of H-PNS for the interaction sites A to D is found to be $0.405e$, $0.14e$, $-0.322e$ and $-0.359e$, respectively. In the case of fluorinated PNS for orientations E to H, the corresponding NBO charges are recorded to be $0.427e$, $0.172e$, $-0.295e$ and $-0.302e$. As a result, it is inferred that the negative value of NBO charge is noticed upon the interaction of NO$_2$ gas on PNS, and the positive value of NBO charge is obtained when the NH$_3$ molecules get adsorbed on PNS \cite{50}. The highest HOMO-LUMO gap changes are obtained for the interaction sites A and E including the high positive magnitude of NBO charge transfer. By contrast, interaction sites B and F have a low positive magnitude of NBO charge transfer with minimum HOMO-LUMO gap changes, even though the corresponding adsorption energy is recorded to be high. In the case of orientations C and G, a low negative magnitude of NBO charge is obtained, which shows a gradual increase in the  average band gap variation. However, high HOMO-LUMO gap deviation is observed for the interaction sites D and H with the low negative magnitude of NBO charge. Thus, it can be concluded that the interaction of hydrogen/oxygen atoms in NH$_3$/NO$_2$ gas molecules on PNS is not favourable. Nevertheless, the electronic properties of phosphorene can be modified with the passivation of hydrogen and fluorine atoms in PNS. Besides, fluorinated PNS shows a significant deviation in the HOMO-LUMO gap, when NH$_3$ or NO$_2$ molecules interact on the phosphorene base material. Moreover, the electronic configuration of fluorine influences the adsorption characteristics of PNS. Figures~\ref{fig-s13}--\ref{fig-s20} refer to the pictorial representation of the HOMO-LUMO gap and DOS-spectrum for the interaction sites A--H. As a result, it is observed from DOS-spectrum that more peak maxima are shifted to LUMO level for PNS. This confirms that the free electrons can easily traverse between NH$_3$/NO$_2$ gas molecules and phosphorene base material, which is an optimum condition for chemical nanosensor.

\begin{figure}[!t]
\centerline{\includegraphics[width=0.65\textwidth]{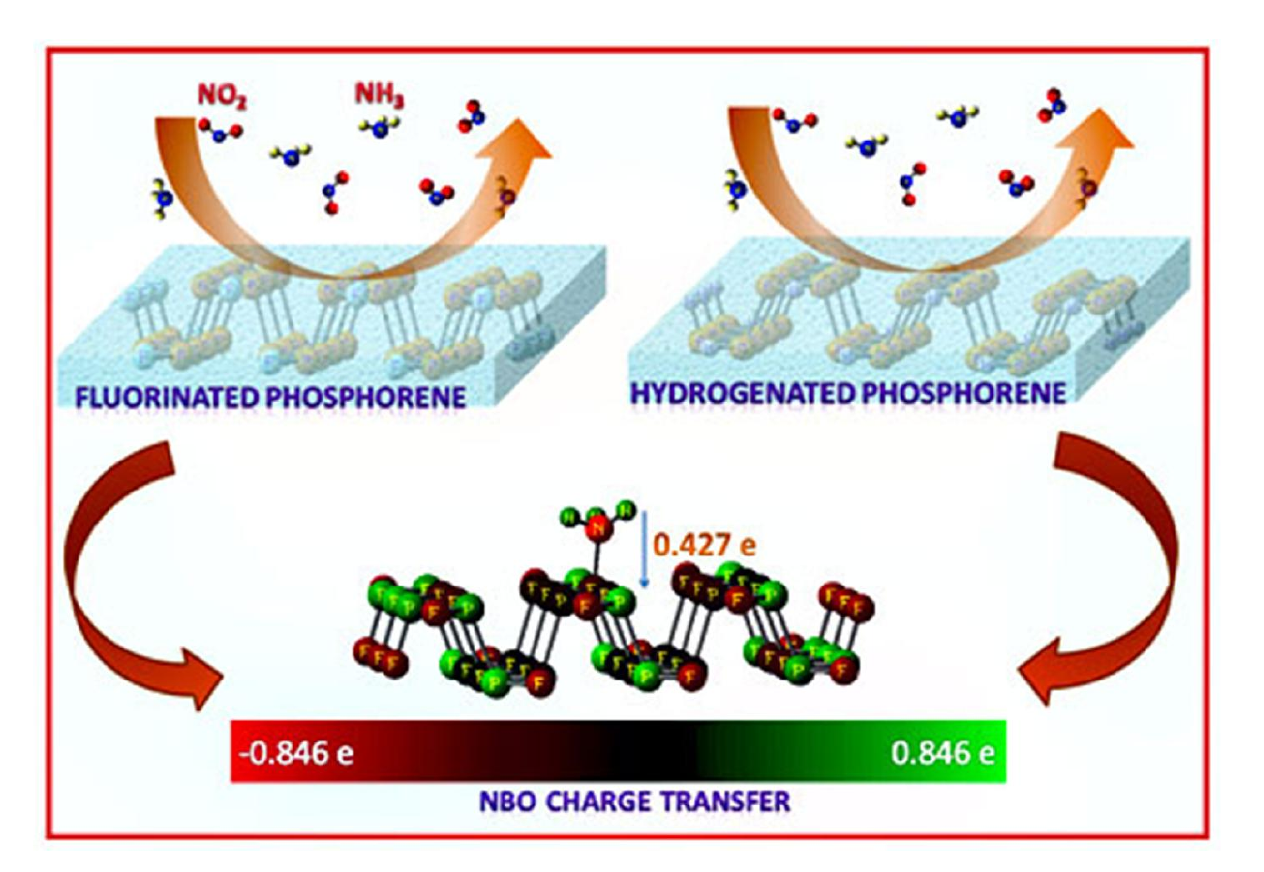}}
\caption{(Colour online) Schematic illustration of NO$_2$/NH$_3$ adsorption on hydrogenated and fluorinated phosphorene nanosheet and NBO charge transfer. } \label{fig-s21}
\end{figure}

Furthermore, alpha ($\alpha$) and beta ($\beta$) orbitals appear in DOS spectrum, which arose owing to the interaction of NO$_2$ gas molecules on both hydrogenated and fluorinated PNS. However, there are no beta or alpha orbitals noticed in isolated PNS. This infers that the interaction of nitrogen dioxide gas molecules on PNS leads to an orbital overlapping of P atoms with NO$_2$ gas, which causes a decrease of HOMO-LUMO gap in PNS. By contrast, the $\alpha$- and $\beta$-orbitals are not observed upon the adsorption of NH$_3$ molecules on PNS. This is also governed by the orbital overlying of NH$_3$ and P atoms. The outcome of DOS-spectrum of H-PNS and F-PNS strongly supports the interaction of nitrogen-based gas molecules on the phosphorene material. Figure~\ref{fig-s21} schematically represents NO$_2$/NH$_3$ adsorption on H-PNS and F-PNS, and NBO charge transfer upon adsorption.

Besides, the suitable interaction site of NH$_3$/NO$_2$ gas molecules on PNS can be determined only after exploring the NBO charge transfer, average HOMO-LUMO gap changes, adsorption energy and HOMO-LUMO gap.

\newpage
\section {Conclusion}
To conclude, the interaction behaviour of NO$_2$ and NH$_3$ molecules on PNS is investigated with the first-principles calculation. The geometric stability of PNS improves with the passivation of hydrogen and fluorine. The stability of the PNS system is verified by the formation energy. The results reveal that the fluorine passivated PNS is relatively more stable than hydrogen passivated PNS. The adsorption of NO$_2$ and NH$_3$ gas molecules on PNS is explored with the significant parameters including NBO charge transfer, adsorption energy, and average HOMO-LUMO gap changes. The strong adsorption sites of these molecules are verified by their electronic properties of the base PNS system upon the interaction of NO$_2$ and NH$_3$. The DOS-spectrum validates the charge transfer between these gas molecules and PNS. The overall results recommend that the PNS can be effectively utilized as a chemical sensor for NO$_2$ and NH$_3$ molecules. 

\section*{Acknowledgements}

The authors wish to express their sincere thanks to Nano Mission Council (No.SR/NM/NS- \linebreak 1011/2017(G)) Department of Science and Technology, India for the financial support.

\ukrainianpart

\title{Вивчення взаємодії молекул газів двоокису азоту та амоніяку на фосфореновому нанолисті --- дослідження методом функціоналу густини }%

\author{В. Нагараджан, Р. Чандірамулі}
\address{Школа електротехнiки та електронiки, Академiя мистецтв, наукових i технологiчних дослiджень Шанмуга
(унiверситет SASTRA), Танджавур, Тамiл-Наду --- 613 401, Iндiя
}

\makeukrtitle

\begin{abstract}
Адсорбційні характеристики небезпечних газових молекул, а саме двоокису вуглецю  (NO$_2$) та амоніяку (NH$_3$), на фосфореновому нанолисті 
досліджено з допомогою  ab initio методики. Для покращення структурної міцності первинного нанолиста,  ми здійснили пасивацію водню і фтору на кінцевому ребрі. 
Структурну міцність пасивованого воднем і фтором фосфоренового листа перевірено з огляду на формування енергії.
Основною метою даної роботи є дослідити гази NO$_2$ і NH$_3$, використовуючи фосфореновий нанолист в якості  базового сенсорного
матеріалу. 
Адсорбцію різноманітних селективних адсобційних вузлів цих газових молекул досліджено у відповідності до змін середньої  HOMO-LUMO щілини, 
перенесення заряду натуральний-зв'язок-орбіталь (NBO),   HOMO-LUMO щілини та енергії адсорбції. Варто відзначити, що знайдено від'ємне значення енергії адсорбції 
після адсорбції  NO$_2$ і NH$_3$ на фосфореновому нанолисті, і воно знаходиться в діапазоні  від
$-1.36$ до $-2.45$~еВ. Результати даної роботи доводять, що гідрогенізований і фторований фосфореновий нанолист можна ефективно використовувати
в якості хімічного сенсора молекул  NO$_2$ і NH$_3$.
\keywords фосфорен, нанолист, адсорбція, NO$_2$, NH$_3$, енергія формування
\end{abstract}


\begin{thebibliography}{99}
\bibitem{1}      Endres H.E., G\"ottler W., Hartinger R., Drost S., Hellmich W., M\"uller G., Bosch-von Braunm\"uhl C., Krenkow~A., Perego C., Sberveglieri G., Sens. Actuators, B, 1996, \textbf{36}, 353, \doi{10.1016/S0925-4005(97)80095-0}.  
\bibitem{2}  	 Tomchenko A.A., Harmer G.P., Marquis B.T., Allen J.W.,  Sens. Actuators, B, 2003, \textbf{93}, 126,\\ \doi{10.1016/S0925-4005(03)00240-5}.
\bibitem{3}	  Seiyama T., Kato A., Fujiishi K., Nagatani M., Anal. Chem., 1962, \textbf{34}, 1502, \doi{10.1021/ac60191a001}.
\bibitem{4}    Arafat M., Dinan B., Akbar S.A., Haseeb A., Sensors, 2012, \textbf{12}, 7207, \doi{10.3390/s120607207}.
\bibitem{5}	Capone S., Forleo A., Francioso L., Rella R., Siciliano P., Spadavecchia J., Presicce D.S., Taurino A.M.,  J.~Optoelectron.  Adv. Mater., 2003, \textbf{5}, 1335.
\bibitem{6}	Nagarajan V., Chandiramouli R., Appl. Surf. Sci., 2017, \textbf{413}, 109, \doi{10.1016/j.apsusc.2017.04.008}. 
\bibitem{7}	Nagarajan V., Chandiramouli R., J. Mol. Graphics Modell., 2018, \textbf{81}, 97, \doi{10.1016/j.jmgm.2018.02.011}. 
\bibitem{8}	Nagarajan V., Chandiramouli R., Appl. Surf. Sci., 2018, \textbf{441}, 734, \doi{10.1016/j.apsusc.2018.02.107}. 
\bibitem{9}      Nagarajan V., Chandiramouli R., Diamond Relat. Mater., 2018, \textbf{85}, 53, \doi{10.1016/j.diamond.2018.03.028}.
\bibitem{10}	Novoselov K.S., Fal'ko V.I., Colombo L., Gellert P.R., Schwab M.G., Kim K., Nature, 2012, \textbf{490}, 192, \doi{10.1038/nature11458}.
\bibitem{11}   Zhang S., Guo S., Chen Z., Wang Y., Gao H., G\'omez-Herrero J., Ares P., Zamora F., Zhu Z., Zeng H., Chem. Soc. Rev., 2018, \textbf{47}, 982, \doi{10.1039/C7CS00125H}.
\bibitem{12}	Kaloni T.P., Schreckenbach G., Freund M.S., J. Phys. Chem. C, 2014, \textbf{118}, 23361, \doi{10.1021/jp505814v}.
\bibitem{13}    Sun Y., Liu Q., Gao S., Cheng H., Lei F., Sun Z., Jiang Y., Su H., Wei S., Xie Y., Nat. Commun., 2013, \textbf{4}, 2899, \doi{10.1038/ncomms3899}.
\bibitem{14}	Lukatskaya M.R., Mashtalir O., Ren C.E., Dall’Agnese Y., Rozier P., Taberna P.L., Naguib M., Simon P., Barsoum M.W., Gogotsi Y., Science, 2013, \textbf{341}, 1502, \doi{10.1126/science.1241488}.
\bibitem{15}	Lei W., Portehault D., Liu D., Qin S., Chen Y., Nat. Commun., 2013, \textbf{4}, 1777, \doi{10.1038/ncomms2818}.
\bibitem{16}	Wu C., Lu X., Peng L., Xu K., Peng X., Huang J., Yu G., Xie Y., Nat. Commun., 2013, \textbf{4}, 2431, \doi{10.1038/ncomms3431}.
\bibitem{17}	Dai J., Zeng X.C., J. Phys. Chem. Lett., 2014, \textbf{5}, 1289, \doi{10.1021/jz500409m}.
\bibitem{18}	Li L., Yu Y., Ye G.J., Ge Q., Ou X., Wu H., Feng D., Chen X.H., Zhang Y., Nat. Nanotechnol., 2014, \textbf{9}, 372, \doi{10.1038/nnano.2014.35}.
\bibitem{19}	Liu H., Neal A.T., Zhu Z., Luo Z., Xu X., Tom\'anek D., Ye P.D., ACS Nano, 2014, \textbf{8}, 4033,\\ \doi{10.1021/nn501226z}.
\bibitem{20}	Xia F., Wang H., Xiao D., Dubey M., Ramasubramaniam A., Nat. Photonics, 2014, \textbf{8},  899, \\ \doi{10.1038/nphoton.2014.271}.
\bibitem{21}	Low J., Cao S., Yu J., Wageh S., Chem. Commun., 2014, \textbf{50}, 10768, \doi{10.1039/C4CC02553A}.
\bibitem{22}	Novoselov K.S., Jiang D., Schedin F., Booth T.J., Khotkevich V.V., Morozov S.V., Geim A.K., Proc. Natl. Acad. Sci. U.S.A., 2005, \textbf{102}, 10451, \doi{10.1073/pnas.0502848102}.
\bibitem{23}	Kou L., Chen C., Smith S.C., J. Phys. Chem. Lett., 2015, \textbf{6}, 2794, \doi{10.1021/acs.jpclett.5b01094}.
\bibitem{24}	Barraza-Lopez S., Kaloni T.P., ACS Cent. Sci., 2018, \textbf{4}, 1436, \doi{10.1021/acscentsci.8b00589}.
\bibitem{25}	Kou L., Frauenheim T., Chen C., J. Phys. Chem. Lett., 2014, \textbf{5}, 2675, \doi{10.1021/jz501188k}.
\bibitem{26}	Lalitha M., Nataraj Y., Lakshmipathi S., Appl. Surf. Sci., 2016, \textbf{377}, 311, \doi{10.1016/j.apsusc.2016.03.119}. 
\bibitem{27}	Ray S.J., Sens. Actuators, B, 2016, \textbf{222}, 492, \doi{10.1016/j.snb.2015.08.039}.
\bibitem{28} 	Cui S., Pu H., Wells S.A., Wen Z., Mao S., Chang J., Hersam M.C., Chen J., Nat. Commun., 2015, \textbf{6}, 8632, \doi{10.1038/ncomms9632}.
\bibitem{29}	 Srivastava A., Khan M.S., Gupta S.K., Pandey R., Appl. Surf. Sci., 2015, \textbf{356}, 881, \\ \doi{10.1016/j.apsusc.2015.08.109}.  
\bibitem{30}	Nagarajan V., Chandiramouli R., J. Mol. Graphics Modell., 2017, \textbf{75}, 365, \doi{10.1016/j.jmgm.2017.06.008}.
\bibitem{31}	Frisch M.J., Trucks G.W., Schlegel H.B., Scuseria G.E., Robb M.A.,  Cheeseman J.R., Scalmani G., Barone V., Mennucci B., Petersson G.A., \textit{et al.}, %Nakatsuji H., Caricato M., Li X., Hratchian H.P., Izmaylov A.F., Bloino J., Zheng G., Sonnenberg J.L., Hada M., Ehara M., Toyota K., Fukuda R., Hasegawa J., Ishida M., Nakajima T., Honda Y., Kitao O., Nakai H., Vreven T., Montgomery J.A. (Jr.), Peralta J.E., Ogliaro F., Bearpark M., Heyd J.J., Brothers E., Kudin K.N., Staroverov V.N., Kobayashi R., Normand J., Raghavachari K., Rendell A., Burant J.C., Iyengar S.S., Tomasi J., Cossi M., Rega N., Millam J.M., Klene M., Knox J.E., Cross J.B., Bakken V., Adamo C., Jaramillo J., Gomperts R., Stratmann R.E., Yazyev O., Austin A.J., Cammi R., Pomelli C., Ochterski J.W., Martin R.L., Morokuma K., Zakrzewski V.G., Voth G.A., Salvador P., Dannenberg J.J., Dapprich S., Daniels A.D., Farkas ?., Foresman J.B., Ortiz J.V., Cioslowski J., Fox D.J., 
Gaussian 09, Gaussian, Inc., Wallingford CT, 2009.
\bibitem{32}	Becke A.D., Phys.  Rev. A, 1988, \textbf{38}, 3098, \doi{10.1103/PhysRevA.38.3098}.
\bibitem{33}	Lee C., Yang W., Parr R.G., Phys.  Rev. B, 1988, \textbf{37}, 785, \doi{10.1103/PhysRevB.37.785}.
\bibitem{34}	Becke A.D., J. Chem. Phys., 1993, \textbf{98}, 1372, \doi{10.1063/1.464304}.
\bibitem{35}	Adamo C., Barone V., J. Chem. Phys., 1999, \textbf{110}, 6158, \doi{10.1063/1.478522}.
\bibitem{36}	Grimme S., Antony J., Ehrlich S., Krieg H., J. Chem. Phys., 2010, \textbf{132}, 154104, \doi{10.1063/1.3382344}.
\bibitem{37}	Kaloni T.P., Giesbrecht P.K., Schreckenbach G., Freund M.S., Chem. Mater., 2017, \textbf{29}, 10248, \\ \doi{10.1021/acs.chemmater.7b03035}.
\bibitem{38}	Barraza-Lopez S., Kaloni T.P., Poudel S.P., Kumar P., Phys. Rev. B, 2018, \textbf{97}, 024110,  \\ \doi{10.1103/PhysRevB.97.024110}.
\bibitem{39}	O'Boyle N.M., Tenderholt A.L., Langner K.M., J. Comput. Chem., 2008, \textbf{29}, 839, \doi{10.1002/jcc.20823}.
\bibitem{40}	Jain A., McGaughey A.J.H., Sci. Rep., 2015, \textbf{5}, 8501, \doi{10.1038/srep08501}.
\bibitem{41}	Ullah H., Shah A.-ul-H.A., Ayub K., Bilal S., J. Phys. Chem. C, 2013, \textbf{117}, 4069, \doi{10.1021/jp311526u}.
\bibitem{42}	Beheshtian J., Peyghan A.A., Noei M., Sens. Actuators, B, 2013, \textbf{181}, 829, \doi{10.1016/j.snb.2013.02.086}.
\bibitem{43}	Boukhvalov D.W., Rudenko A.N., Prishchenko D.A., Mazurenko V.G., Katsnelson M.I., Phys. Chem. Chem. Phys., 2015, \textbf{17}, 15209, \doi{10.1039/C5CP01901J}.
\bibitem{44}	Pinaud B.A., Benck J.D., Seitz L.C., Forman A.J., Chen Z., Deutsch T.G., James B.D., Baum K.N., Baum G.N., Ardo S., Wang H., Miller E., Jaramillo T.F., Energy Environ. Sci., 2013, \textbf{6}, 1983, \doi{10.1039/c3ee40831k}.
\bibitem{45} 	Wang X., Jones A.M., Seyler K.L., Tran V., Jia Y., Zhao H., Wang H., Yang L., Xu X., Xia F., Nat. Nanotechnol., 2015, \textbf{10}, 517, \doi{10.1038/nnano.2015.71}.
\bibitem{46}	Qiao J., Kong X., Hu Z.-X., Yang F., Ji W., Nat. Commun., 2014, \textbf{5}, 4475, \doi{10.1038/ncomms5475}.
\bibitem{47}	Ezawa M., New J. Phys., 2014, \textbf{16}, 115004, \doi{10.1088/1367-2630/16/11/115004}.  
\bibitem{48}	Srimathi U., Nagarajan V., Chandiramouli R., Comput. Theor. Chem., 2018, \textbf{1130}, 68,  \\ \doi{10.1016/j.comptc.2018.03.011}. 
\bibitem{49}	Hu W., Lin L., Yang C., Dai J., Yang J., Nano Lett., 2016, \textbf{16}, 1675, \doi{10.1021/acs.nanolett.5b04593}.
\bibitem{50}	Cai Y., Ke Q., Zhang G., Zhang Y.-W., J. Phys. Chem. C, 2015, \textbf{119}, 3102, \doi{10.1021/jp510863p}.
\bibitem{51}	Mulliken R.S., J. Chem. Phys., 1955, \textbf{23}, 1833, \doi{10.1063/1.1740588}.
\bibitem{52} 	Rastegar S.F., Peyghan A.A., Hadipour N.L., Appl. Surf. Sci., 2013, \textbf{265}, 412, \doi{10.1016/j.apsusc.2012.11.021}.
\bibitem{53}	Prasongkit J., Feliciano G.T., Rocha A.R., He Y., Osotchan T., Ahuja R., Scheicher R.H., Sci. Rep., 2015, \textbf{5}, 17560, \doi{10.1038/srep17560}.
\bibitem{54} 	Nagarajan V., Chandiramouli R., Chem. Phys. Lett., 2018, \textbf{695}, 162, \doi{10.1016/j.cplett.2018.02.019}.
\bibitem{55} 	Nagarajan V., Chandiramouli R., J. Mol. Liq., 2018, \textbf{249}, 24, \doi{10.1016/j.molliq.2017.11.007}.
\bibitem{56}   Nagarajan V., Chandiramouli R., Mater. Sci. Eng., B, 2018, \textbf{229}, 193, \doi{10.1016/j.mseb.2017.12.015}.

\end{thebibliography}
\end{document}